\documentclass[fleqn,usenatbib]{mnras}

\usepackage{newtxtext,newtxmath}
\usepackage{xcolor}
\usepackage{placeins}
\usepackage{booktabs}\usepackage{float}

\usepackage[T1]{fontenc}

\DeclareRobustCommand{\VAN}[3]{#2}
\let\VANthebibliography\thebibliography
\def\thebibliography{\DeclareRobustCommand{\VAN}[3]{##3}\VANthebibliography}

\usepackage{graphicx}	
\usepackage{amsmath}	
\usepackage{multicol}
\usepackage{multirow}

\title[Unresolved r and g modes in 214 A and F stars]{Unresolved Rossby and gravity modes in 214 A and F stars showing rotational modulation}

\author[A. I. Henriksen et al.]{Andreea I. Henriksen,$^{1}$\thanks{E-mail: andreea@space.dtu.dk}
Victoria Antoci,$^{1}$\thanks{E-mail: antoci@space.dtu.dk}
Hideyuki Saio,$^{2}$
Frank Grundahl, $^{3}$
Hans Kjeldsen, $^{3}$ 
\newauthor
Timothy Van Reeth, $^{4}$
Dominic M. Bowman, $^{4}$
Péter I. Pápics, $^{4}$
Peter De Cat, $^{5}$
Joachim Kr\"uger, $^{6,7}$
\newauthor
M. Fredslund Andersen,  $^{3}$
P. L. Pall\'e $^{8,9}$
\\
$^{1}$National Space Institute, Technical University of Denmark, Elektrovej, DK-2800 Kgs. Lyngby, Denmark\\
$^{2}$Astronomical Institute, Graduate School of Science, Tohoku University, Sendai 980-8578, Japan\\
$^{3}$ Stellar Astrophysics Centre, Department of Physics and Astronomy, Aarhus University, DK-8000 Aarhus C, Denmark \\
$^{4}$ Institute of Astronomy, KU Leuven, Celestijnenlaan 200D, 3001 Leuven, Belgium \\
$^{5}$ Royal Observatory of Belgium, Ringlaan 3, 1180 Brussels, Belgium \\
$^{6}$ Centre for Astrophysics, University of Southern Queensland, Toowoomba, QLD 4350, Australia\\
$^{7}$ Astronomical Observatory Institute, Faculty of Physics, A.Mickiewicz University, ul. S\l{}oneczna 36, 60-286 Poznan, Poland\\
$^{8}$ Instituto de Astrofísica de Canarias, 38205 La Laguna, Tenerife, Spain \\
$^{9}$ Departamento de Astrofísica, Universidad de La Laguna (ULL), 38206 La Laguna, Tenerife, Spain \\
}

\date{Accepted XXX. Received YYY; in original form ZZZ}

\pubyear{20xx}

\begin{document}
\label{firstpage}
\pagerange{\pageref{firstpage}--\pageref{lastpage}}
\maketitle

\begin{abstract} 
Here we report an ensemble study of 214 A- and F-type stars observed by \textit{Kepler}, exhibiting the so-called \textit{hump and spike} periodic signal, explained by Rossby modes (r~modes) - the \textit{hump} - and magnetic stellar spots or overstable convective (OsC) modes- the \textit{spike}, respectively. We determine the power confined in the non-resolved hump features and find additional gravity~modes (g~modes) humps always occurring at higher frequencies than the spike. Furthermore, we derive projected rotational velocities from FIES, SONG and HERMES spectra for 28 stars and the stellar inclination angle for 89 stars. We find a strong correlation between the spike amplitude and the power in the r and g~modes, which suggests that both types of oscillations are mechanically excited by either stellar spots or OsC modes. Our analysis suggests that stars with a higher power in $m=1$ r~modes humps are more likely to also exhibit humps at higher azimuthal orders ($m$ = 2, 3, or 4). Interestingly, all stars that show g~modes humps are hotter and more luminous than the observed red edge of the $\delta$ Scuti instability strip, suggesting that either magnetic fields or convection in the outer layers could play an important role.

\end{abstract}




\section{Introduction}
Rossby or r modes \citep{Rossby1939RelationBV, 1978MNRAS.182..423P}, are retrograde toroidal motions, which in rotating stars, couple with spheroidal motions (restored by the Coriolis force) and cause temperature perturbations, generating observable effects \citep{2018MNRAS.474.2774S}. Out of these, the toroidal component is responsible for most of the matter flow patterns on the stellar surface, while the spheroidal component causes smaller displacements.

Rossby~modes have been the subject of various other astrophysical theoretical studies (e.g.,
\citealt{Provost1981}, \citealt{1982ApJ...256..717S},  \citealt{1998ApJ...502..708A}), but it was not until photometric space missions, such as \textit{Kepler} \citep{2010ApJ...713L..79K} or TESS \citep{2015JATIS...1a4003R}, that  r~modes could be detected in stars. Initially reported in a sample of $\gamma$ Doradus stars \citep{2016A&A...593A.120V}, it was later suggested that r~modes could be present in other types of stars \citep{2018MNRAS.474.2774S, 2018phos.confE..41S}. From hot subdwarfs \citep{2020MNRAS.496..718J} to intermediate-mass stars (e.g., \citealt{2016A&A...593A.120V}, \citealt{2018MNRAS.474.2774S}, \citealt{2019MNRAS.487..782L}) it appears that the latest advancement in the observational effort, has revealed that r~modes are seemingly common in early type stars across their evolution.

\citet{2018NatAs...2..568L} have shown the presence and importance of r~modes in the shallow near surface layers of our Sun, arguing that, as the vorticity of these waves is comparable to that of the convection, they play an essential role in solar dynamics.

Studying the signal from r~modes can also prove useful in exoplanet and stellar activity studies. \citet{2019A&A...623A..50L} investigated the radial velocity variation that sectoral r~modes would induce in the case of Solar-like stars and concluded that their signal could lead to inaccurate results if not taken into account. In addition, r~modes can prove useful in studying binary systems as \citet{2018MNRAS.474.2774S} reported the possibility of r~modes to be generated by tidal disturbances in so-called heartbeat stars. A later study on 20 such tidally interacting stars \citep{2022MNRAS.511..560S} showed that by fitting the expected visibility distribution of r~modes, the stellar rotation rates could be determined. 

In the case of intermediate-mass stars, analysing r~modes together with gravity (g)~modes helps to better understand and determine the internal stellar rotation rate \citep{2019MNRAS.487..782L,2020A&A...644A.138T}.
Furthermore, \citet{2018MNRAS.474.2774S} argued that r~modes could be mechanically excited by the disruption of the rotational flow on the stellar surface caused by stellar spots. This phenomenon would give rise to a specific periodic signal, \textit{hump and spike}, found in hundreds of early-type stars (e.g., \citealt{2013MNRAS.431.2240B, 2015MNRAS.448.1378B, 2020MNRAS.492.3143T,Henriksen23}), where the \textit{hump} represents unresolved r~modes and the spike represents the spots that are co-rotating with the stellar surface. 

\citet{Henriksen23} showed that the spike, if due to stellar spots, could be translated into an estimate of the strength of the magnetic field, as the spike amplitudes are anti-correlated with the stellar mass, which agrees with the theoretical predictions from \cite{2019ApJ...883..106C}. Furthermore, it appears that the lifetimes of the spike are short, as expected from magnetic features of subsurface dynamo generate fields. On the other hand, the rotational modulation that is translated into the spike could also be attributed to overstable convective (OsC) modes that resonantly excite low-frequency g~modes in the stellar envelope \citep{LeeSaio2020, 2021MNRAS.505.1495L}. While the harmonic signature of around 20 per cent of the \textit{hump and spike} stars studied in \citealt{Henriksen23}, indicate a signal consistent with stellar spots, for the larger part of the stars in the sample, the OsC modes cannot be ruled as the cause of \textit{spike} feature. 
This calls for additional analysis to understand the phenomenon behind the \textit{spike} feature, for example, direct magnetic field measurements. 

Another approach is to study the \textit{hump} feature since the r~modes that are causing it are suggested to be connected to the phenomenon that gives rise to the \textit{spike}. In contrast to the r~modes identified in previous studies of intermediate-mass stars (e.g.,\citealt{2016A&A...593A.120V, 2019MNRAS.487..782L,2020A&A...644A.138T}), the r~modes hypothesized to be present in \textit{hump and spike} stars, have very high radial orders. This means that, despite the 4-yr \textit{Kepler} data set, these modes are unresolved and a period spacing cannot be determined \citep{2018MNRAS.474.2774S}. Here we focus on finding correlations between the characteristics of the \textit{hump} and the \textit{spike}. The sample, data and analysis methods are described in Section \ref{sec:data_analysis}. We devote section \ref{sec:result} for discussing and interpreting our results. We conclude our work in Section \ref{sec:conc}.

\section{Data analysis}
\label{sec:data_analysis}

\subsection{Sample selection}
\label{sec:sample}

\begin{figure}
	\includegraphics[width=\columnwidth]{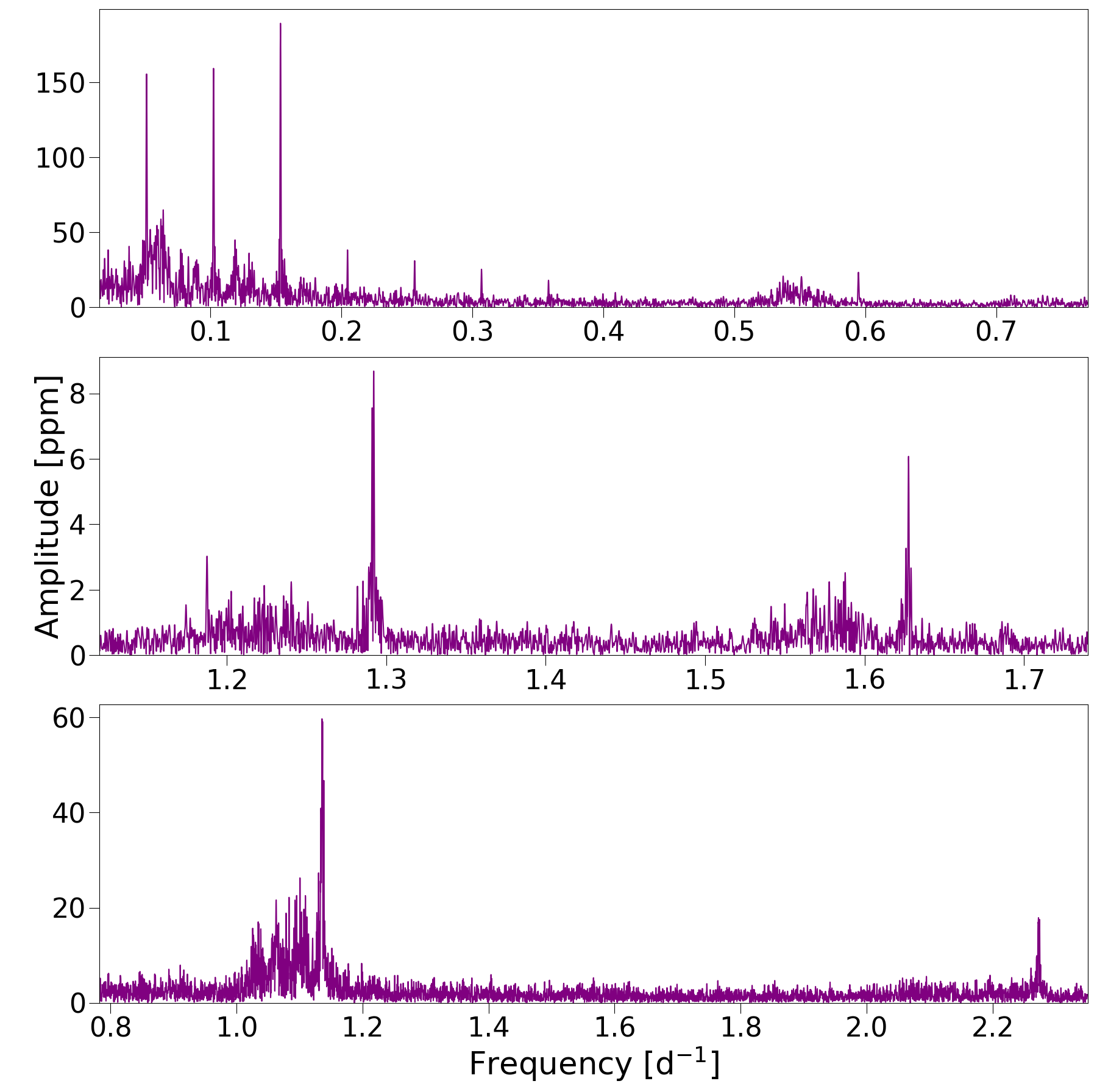}
    \caption{Fourier spectra of two targets from our sample that show signs of being in a binary system and of Tabby's star. \textit{Upper panel:} Fourier spectrum of KIC\,6116612. The \textit{hump and spike} feature is located between  $\sim 0.5 - 0.6\,\rm d^{-1}$, while at lower frequencies ($\leq 0.4~\rm d^{-1}$), clear periodic signal of a transiting companion. \textit{Middle panel:} Fourier spectrum of KIC\,5980337. The \textit{hump and spike} feature appears twice in the spectrum. Given that the frequency of the second spike is not a harmonic of the first spike, it could indicate that the Fourier spectrum is a composite spectrum of two \textit{hump and spike} stars (possibly  in a binary system). \textit{Lower panel:} Fourier spectrum of KIC\,8462852 (Tabby's star) reveals a hump between $\sim 1.01$ and $1.12\,{\rm d^{-1}}$ and a spike at $\sim 1.136\,{\rm d^{-1}}$, as well as the second harmonic of the spike at $\sim\,2.27\,{\rm d^{-1}}$. The third harmonic is detected as well ($f=3.409\,{\rm d^{-1}}$).}
    \label{fig:example_hump_spike}
\end{figure}

In \citealt{Henriksen23}, 162 stars that exhibit the \textit{hump and spike} feature in their Fourier spectra computed with \textit{Kepler} data were studied. In addition to these stars, we include 52 previously discarded stars in our sample. \citealt{Henriksen23} did not analyse these targets as they showed possible signs of binarity, which would alter stellar parameters, such as $T_{\rm eff}$ or luminosity. Fig. \ref{fig:example_hump_spike} depicts some examples of such stars. We chose to include these stars in our sample, as we will search for correlation between parameters of the \textit{hump and spike} feature, which are determined independently of stellar parameters. As mentioned in \citealt{Henriksen23}, the targets were selected from literature, from works such as  \cite{2013MNRAS.431.2240B, 2014MNRAS.441.3543B, 2017MNRAS.467.1830B}, \citealt{2015MNRAS.448.1378B}, \citet{2020MNRAS.492.3143T} and \citet{2021ApJS..255...17S} or by visually inspecting the Fourier spectra of \textit{Kepler} time series. In total, we focus on studying 214 \textit{hump and spike} stars by analysing photometric and spectroscopic data to shed light on the phenomenon that could cause the \textit{hump} feature.

\subsection{Photometry}
\label{sec:photo}

\subsubsection{Kepler light curves}

\begin{figure}
	\includegraphics[width=\columnwidth]{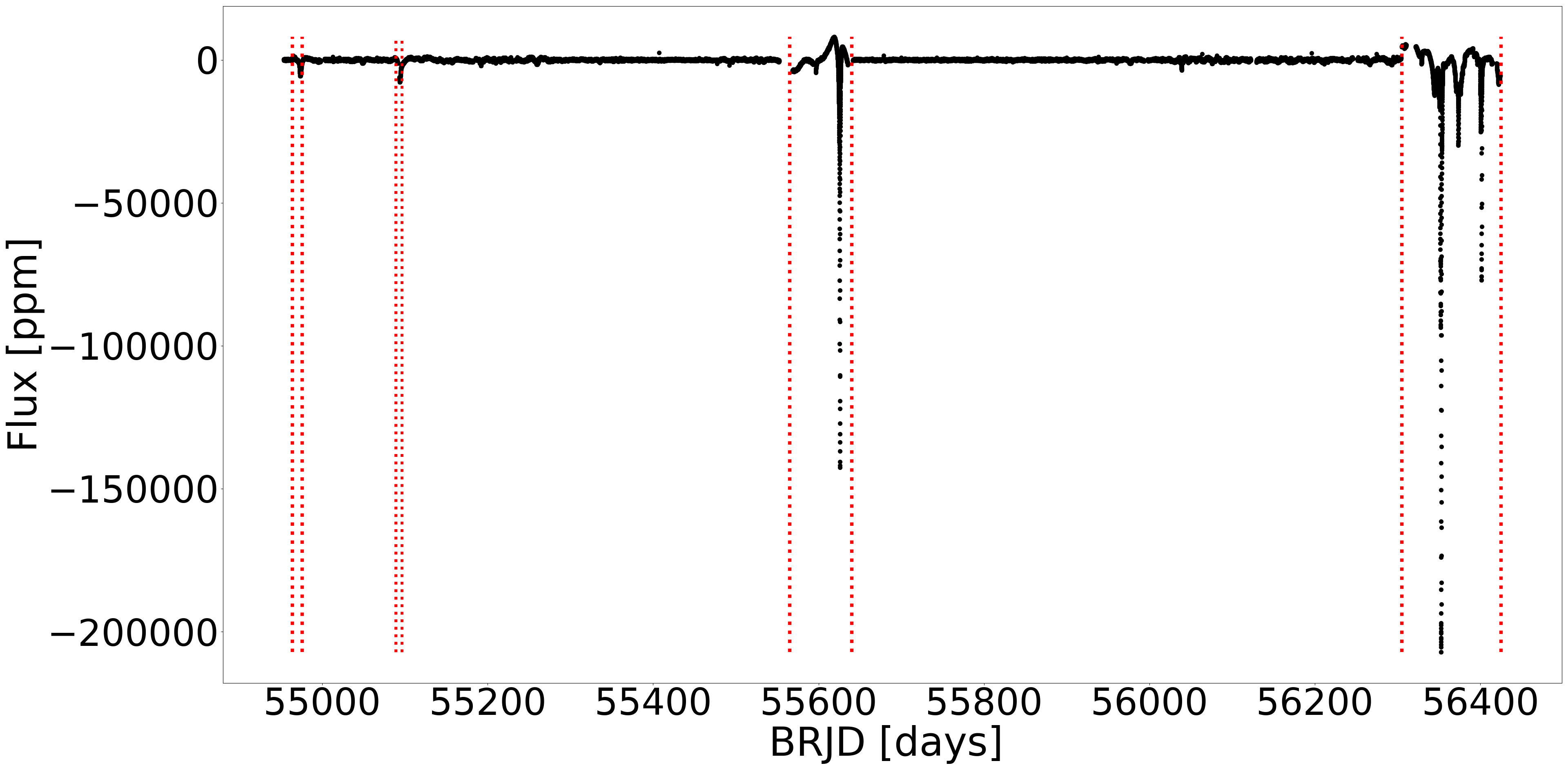}
    \caption{Lightcurve of KIC\,8462852 (Tabby's star). The vertical red lines indicate the data that were excluded when computing the Fourier spectrum from lower panel of Fig. \ref{fig:example_hump_spike}.} \label{fig:KIC8462852_lightcurve}
\end{figure}

As in \citealt{Henriksen23}, we have made use of the \textit{Kepler} long cadence data \citep{2010ApJ...713L..79K}, which were retrieved from KASOC (Kepler Asteroseismic Science Operations Center) \footnote[1]{\url{kasoc.phys.au.dk}}. The PDC (Pre-search Data Conditioning) data products were used in our analysis as they did not contain specific systematic errors and the \textit{hump} signal was unaltered by the Kepler Science Operations Center Pipeline \citep{2010SPIE.7740E..1UT}. Based on the quarterly stitched data, a Fourier spectrum was computed for each star, with the help of the \textit{lightkurve} python package \citep{2018ascl.soft12013L}.
KIC\,8462852 (Tabby's star, \citealt{2016MNRAS.457.3988B}) was included in our sample, and due to its challenging lightcurve (see Fig. \ref{fig:KIC8462852_lightcurve}), and rather than correcting the irregularly deep dips found in the time series, we chose to exclude parts of it. The timestamps (in BRJD - Barycentric Reduced Julian Date = BJD-2,400,000.0) of the excluded data, indicated with red vertical lines in Fig. \ref{fig:KIC8462852_lightcurve} are as follows: between 54964 and 54976, between 55089 and 55096, between 55565 and 55640, and everything after 56305. The data in these time intervals correspond to the most significant dips in the light curve (dips 1, 2, 5, 7-10 from \citealt{2016MNRAS.457.3988B}). For the scope of our work, we deemed that excluding the data between these time ranges was sufficient, as this revealed the \textit{hump and spike} signal in the Fourier spectrum (identified first by \citealt{2016MNRAS.457.3988B}) and allowed further analysis. For more information regarding the analysis of photometric data for the remaining 213 stars in our sample, the reader is referred to \citealt{Henriksen23}. 

\subsubsection{Characterizing the hump}
\label{sec:hump_char}
\begin{figure}
	\includegraphics[width=\columnwidth]{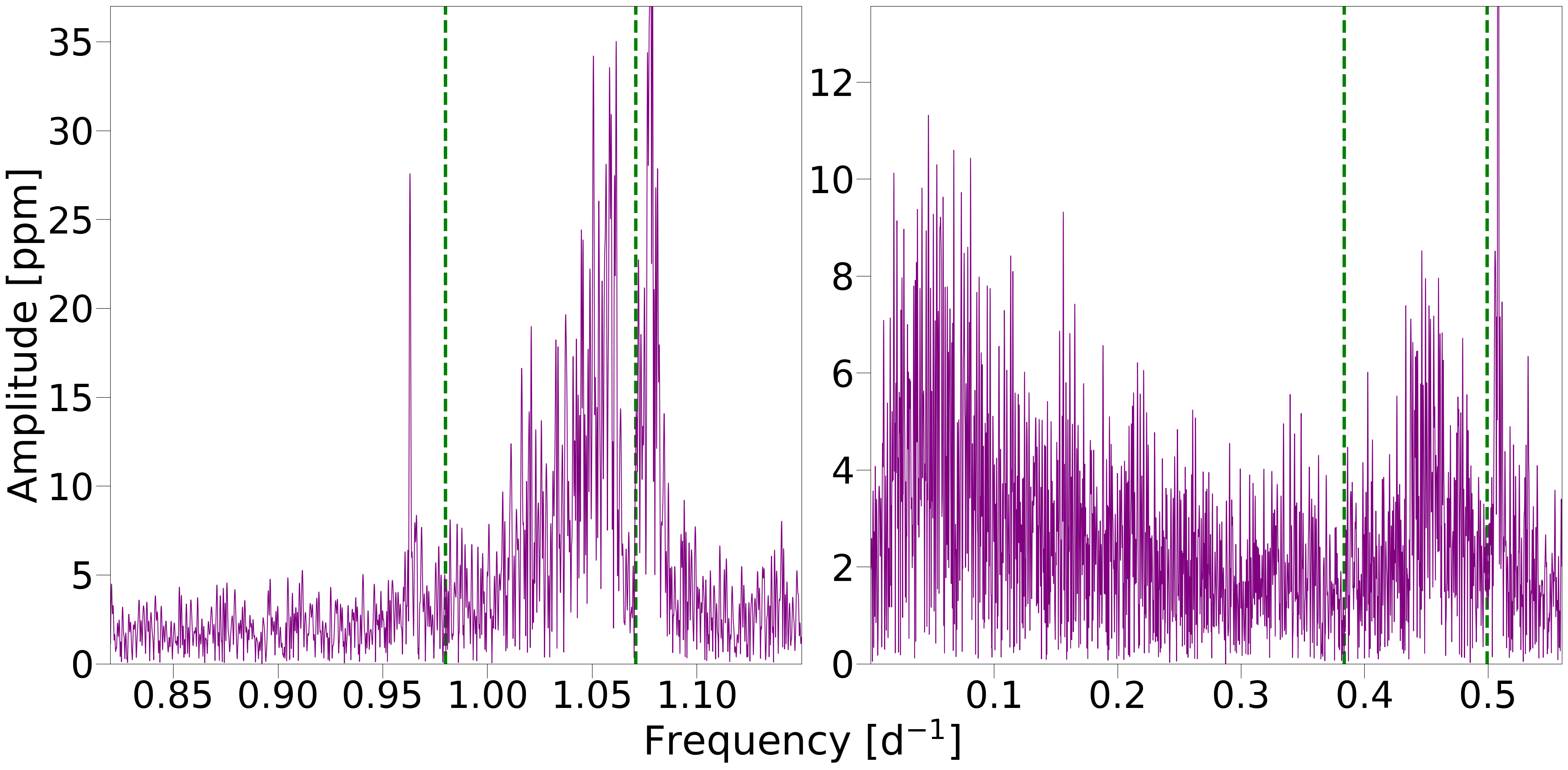}
    \caption{Two targets from our sample for which a manual selection of the hump frequency region was necessary. For a better visualisation of the humps, both panels are focused on lower SNR signals than the spike, which causes the spikes to be clipped. Vertical green lines indicated the selected frequency limits. \textit{Left panel}: KIC\,5632439. Spike: frequency = $1.078\,\rm {d^{-1}}$; amplitude = $49\,\rm{ppm}$. \textit{Right panel}: KIC\,10548172. Spike frequency = $0.508\,\rm {d^{-1}}$; amplitude = $30\,\rm{ppm}$. }
    \label{fig:example_hump_selection_difficult}
\end{figure}

\begin{figure}

	\includegraphics[width=\columnwidth]{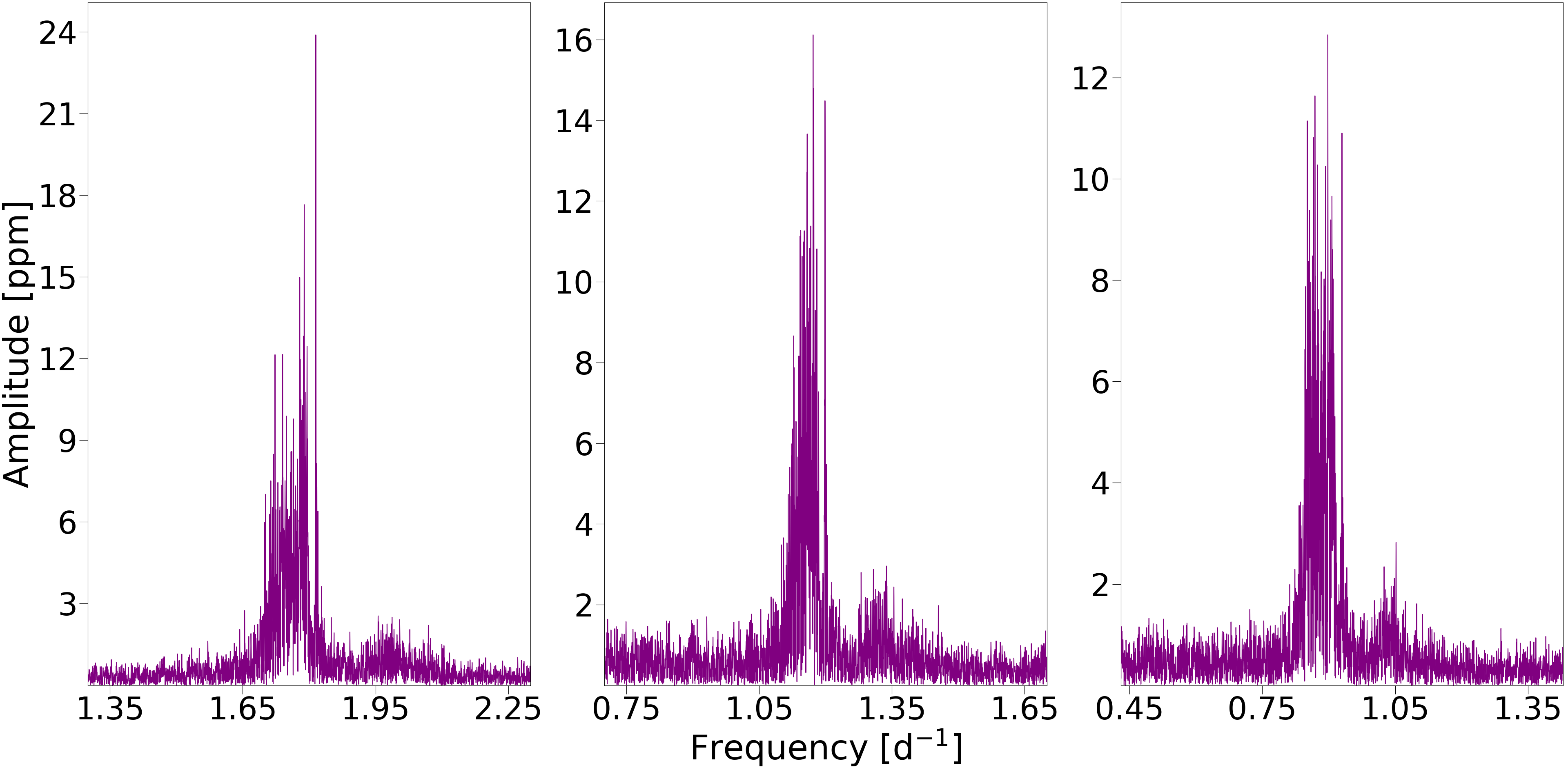}
    \caption{Example of three targets exhibiting the \textit{hump after spike} (g modes hump) feature. \textit{Left panel}: KIC\,5524045. \textit{Middle panel}: KIC\,7842339. \textit{Right panel}: KIC\,9273647.} 
    \label{fig:example_hump_after_spike}
\end{figure}

The hump is always a broad feature proximate to a sharp \textit{spike}, found at the high-frequency end of the hump. However, the \textit{hump} feature differs from star to star in terms of amplitude and width. For example, in some stars, the amplitude of the highest peaks is comparable to that of the spike (upper panel of Fig. \ref{fig:example_hump_spike}, left panel of Fig. \ref{fig:example_hump_selection_difficult}) or even higher (see more examples in fig. 1 in \citealt{Henriksen23}). In other cases, the hump is significantly lower in amplitude than the spike (lower panel of Fig. \ref{fig:example_hump_spike}, right panel of Fig. \ref{fig:example_hump_selection_difficult}; additional examples can be found in fig. 1 in \citealt{Henriksen23}). Many stars exhibit spike harmonics: 194 stars (90 per cent) with second harmonic, 115 stars (53 per cent) with third harmonic, and 69 stars (32 per cent) with fourth harmonic. Occasionally, we observed also humps located before two, three or four times the main spike frequency (116 stars with a hump before the second spike harmonic [54 per cent], 14 stars with a hump before the third spike harmonic [6 per cent], four stars with a hump before the fourth spike harmonic [two per cent]). The hump before the main spike corresponds to $m=1$ even r~modes, while the humps located before the harmonics are $m=2$, $m=3$, and $m=4$ r~modes, respectively.
Another feature of interest in this work, identified in 84 stars (39 per cent), is a similar, broad, unresolved hump visible at slightly higher frequencies than the main spike (see examples in Fig. \ref{fig:example_hump_after_spike}). \citet{2018MNRAS.474.2774S} suggested that this low-amplitude feature is a group of prograde dipole g~modes. 
In order to make a distinction between the two types of hump, we refer to the hump located at lower frequencies with respect to the spike as r~modes hump and the one at higher frequencies as g~modes hump. 

The hump signal was identified in each Fourier spectrum and the frequency range where a hump lies was selected manually for each star. The reason why a manual selection was preferred was due to the large diversity in the hump variability from target to target, and the occasional presence of nearby peaks that should not be considered part of the hump, making automatic selection unfeasible. Two such examples are given in Fig. \ref{fig:example_hump_selection_difficult}. In the left panel, at around $\rm 0.95\,d^{-1}$, KIC\,5632439 exhibits a high amplitude peak which is unlikely to be due to an r~mode given the theoretical predicted visibilities \citep{2018MNRAS.474.2774S}. The proximity to the hump would have caused issues in the case of an automatic selection. However, we cannot exclude that the peak is an r~mode excited by a different mechanism. In the right panel of Fig. \ref{fig:example_hump_selection_difficult}, the r~modes hump of KIC\,10548172 is located close to low-frequency peaks that are likely instrumental or due to the presence of a companion. Regardless of the origin of this signal, the proximity of the hump to these peaks, in the case of an automatic selection of the hump limits, would have caused a wider frequency range selection of the hump.

The accuracy with which we have selected the hump has been tested on synthetic data. For this, we have used the frequencies and visibility of $m=1$ even r~modes models (with rotation frequencies of $f_{\rm rot}= 1\,{\rm d^{-1}}$ and  $f_{\rm rot}= 2\,{\rm d^{-1}}$, inclination = 90 deg for both models), described and depicted in fig. 4 in \citet{2018MNRAS.474.2774S}. A synthetic time series was computed and Gaussian noise was added. Different SNRs were used, which are computed as the ratio between the highest peak and the noise level. As seen in Table \ref{tab:synthetic_hump_freq}, as the noise level increases, the low-frequency limit of the hump is less accurate.  However, even in the case of no added noise, the low-frequency part of the hump cannot be recovered because of the low amplitudes of the low-frequency r~modes. The high-frequency limit of the hump is recovered with more confidence because of the higher expected visibility of r~modes at higher frequencies. Additionally, an observed hump is also delimited by the spike, which aids in the selection, as the frequencies of r~modes ($m=1$) are always lower than the rotation frequency ($f_{\rm rm}<f_{\rm rot}$) in the inertial frame \citep{2018MNRAS.474.2774S}.

\subsubsection{r~modes detectability}
\label{sec:r_mode_detect}
\begin{figure}
	\includegraphics[width=\columnwidth]{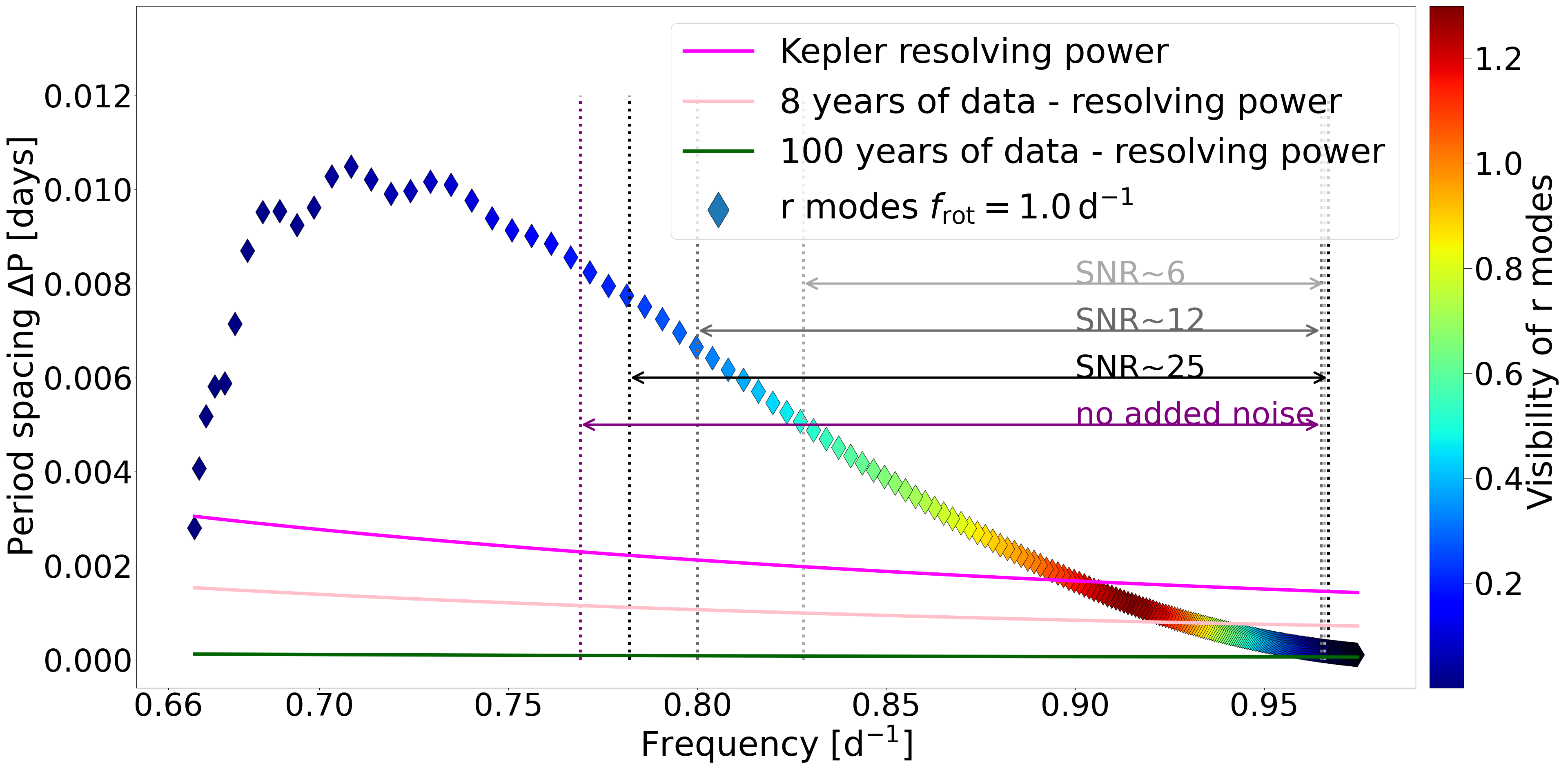}
    \includegraphics[width=\columnwidth]{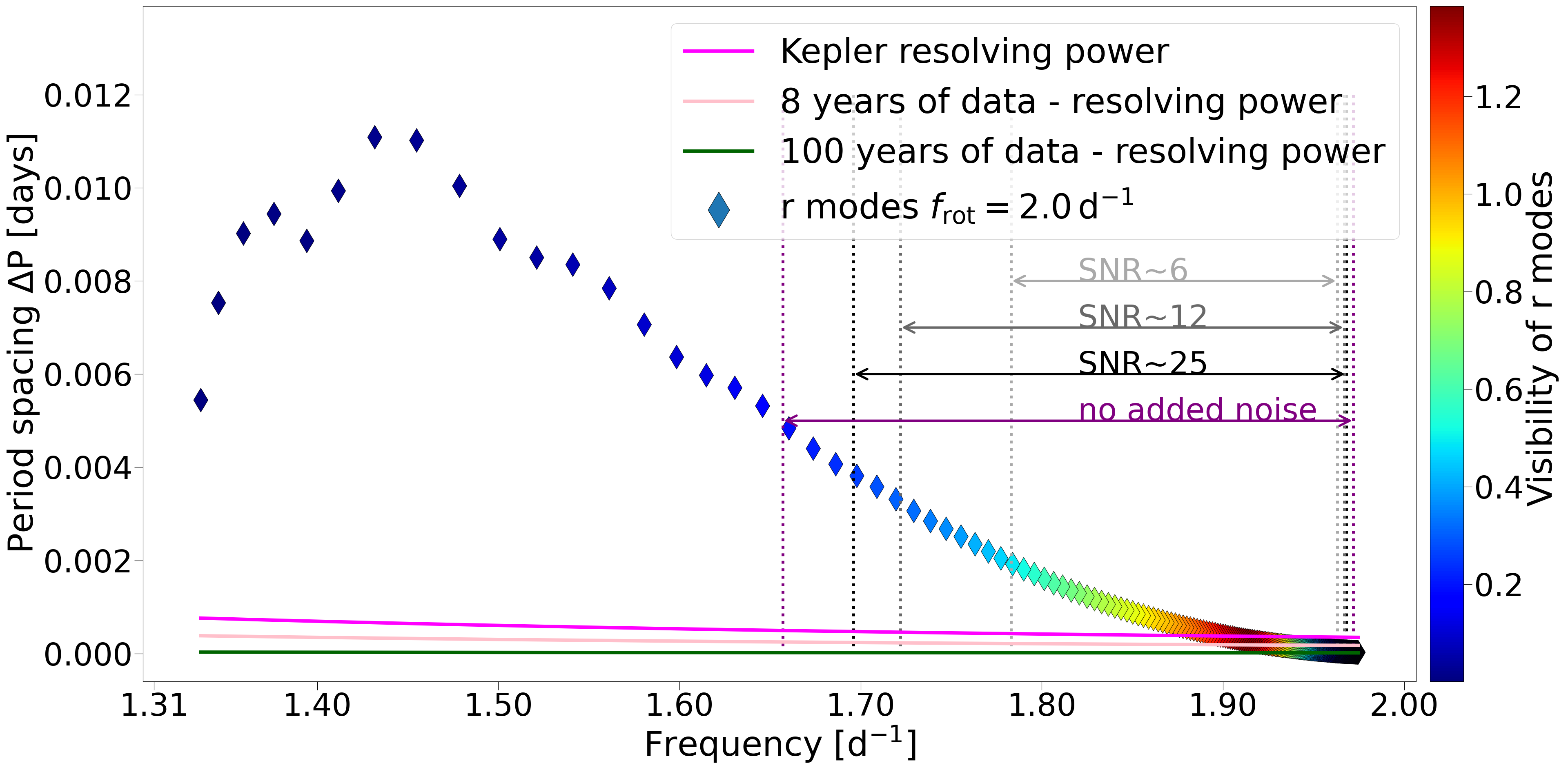}
    \caption{Period spacings of theoretical r~modes based on models from \citet{2018MNRAS.474.2774S} (see text for more details), $(m, k) =(1, -2)$ in a $1.6~{\rm M_{\odot}}$, $T_{\rm eff}= 7150\,{\rm K}$, $1.89~{\rm R_{\odot}}$ model. The hump widths as described in Table \ref{tab:synthetic_hump_freq}  in the case of no added noise and a hump with SNR = 25, 12 and 6, respectively are delimited by the horizontal double sided arrows labeled and accordingly. The \textit{Kepler} resolving power is highlighted in magenta, while the resolving powers in the case of longer time series is colored in pink (8 years of data) and green (100 years of data). \textit{Upper panel}: $f_{\rm rot}= 1.0\,{\rm d}^{-1}$. \textit{Lower panel}: $f_{\rm rot}= 2.0\,{\rm d}^{-1}$.}
    \label{fig:period_spacing_hump}
\end{figure}

The r~modes identified in previous studies of intermediate-mass stars 
(\citealt{2016A&A...593A.120V, 2019MNRAS.487..782L,2020A&A...644A.138T}) were resolved, and so a period spacing could be determined. In our case, however, as \citet{2018MNRAS.474.2774S} pointed out, r~modes cannot be resolved despite the 4-yr \textit{Kepler} data set. This is because of their very high radial orders and small period spacings, as illustrated in Fig. \ref{fig:period_spacing_hump}, where we depict the same two synthetic models mentioned in section \ref{sec:hump_char}. While the period spacing at low frequencies is large enough to be detected with the \textit{Kepler} resolving power, the visibility of the r~modes at lower frequencies is too low for them to be identified (see the color bar in Fig. \ref{fig:period_spacing_hump}). In Fig. \ref{fig:period_spacing_hump}, the vertical lines, annotated with the double-sided arrows, delimit the selected frequency interval from Table \ref{tab:synthetic_hump_freq}. Furthermore, despite the r~modes peaks being visible at higher frequencies, the period spacing becomes too small to be detectable with the \textit{Kepler} data set. Not even with a time series twice as long would all the r~modes in the hump be resolved. For the two cases displayed here, a time series of 100~yr would be necessary in order to resolve all the r modes. This indicates that the widths of the selected humps in this work are underestimated at the lower-frequency limit and that, in fact, r~modes span over wider frequency ranges. The selected width of the theoretical r~modes humps span from $0.15$ to $0.3\,{\rm d^{-1}}$ (depending on the rotation frequency or noise level (see Table \ref{tab:synthetic_hump_freq} and Fig. \ref{fig:period_spacing_hump}). This is in agreement with the selected widths of the observed humps of our stars. The average hump width in our sample is $\sim 0.1 \,{\rm d^{-1}}$, while the maximum and minimum hump width is $\sim\,0.21\,{\rm d}^{-1}$ and $\sim\,0.022\,{\rm d}^{-1}$, respectively. For around 70 per cent of our stars, the SNR of the r~modes hump is lower than 30. 

\begin{table}
    \centering
	\caption{Selection of frequency limits of two theoretical r~modes humps from models with rotation frequencies of $1.0 \rm\,d^{-1}$ and $2.0\,\rm d^{-1}$, respectively. ${\rm f_{left}}$ and ${\rm f_{right}}$ represent the frequency limits of the humps at lower and higher frequencies, respectively. Input values are theoretical values from the model depicted in fig. 4 in \citet{2018MNRAS.474.2774S}. The selected values are average values obtained from 10 realizations. The accuracy of the selection is written in the parenthesis following each value.}
 
    \setlength{\tabcolsep}{1.7pt} 
	
	\label{tab:synthetic_hump_freq}
	\begin{tabular}{|c|cc|cc|} 
    \hline
    Noise &\multicolumn{2}{c}{${\rm f_{left}} [{\rm d^{-1}]}$} &\multicolumn{2}{c}{${\rm f_{right}} [{\rm d^{-1}]}$}\\
     \cline{2-5}
    level& Input & Selected  & Input & Selected \\
     \hline
        & & $f_{\rm rot} = 1.0 {\rm\,d^{-1}}$ &&  \\
     \hline
    no noise & 0.667  & $0.769\,(0.001)$ &0.975 &$0.965\,(0.001)$ \\
    SNR$\sim 25$ & & $0.782\,(0.011)$ & & $0.967\,(0.003)$\\
     
    SNR$\sim 12$ &  & $0.800\,(0.011)$ &  &$0.965\,(0.003)$ \\
     
    SNR$\sim 6$ &  & $0.828\,(0.018)$ &  &$0.966\,(0.009)$ \\

    \hline
    & & $f_{\rm rot} = 2.0 {\rm\,d^{-1}}$ &&  \\
    \hline
    no noise & 1.336 & $1.657\,(0.001)$ & 1.975 &$1.972\,(0.004)$ \\
    SNR$\sim 25$ & & $1.696\,(0.016)$ &  &$1.968\,(0.005)$ \\
    SNR$\sim 12$ &  & $1.722\,(0.039)$ &  &$1.967\,(0.006)$\\
    SNR$\sim 6$&  & $1.783\,(0.024)$ &  &$1.963\,(0.009)$ \\

    \hline
    \hline
    \end{tabular}
\end{table}

\subsubsection{The power of the humps}
\begin{figure}
	\includegraphics[width=\columnwidth]{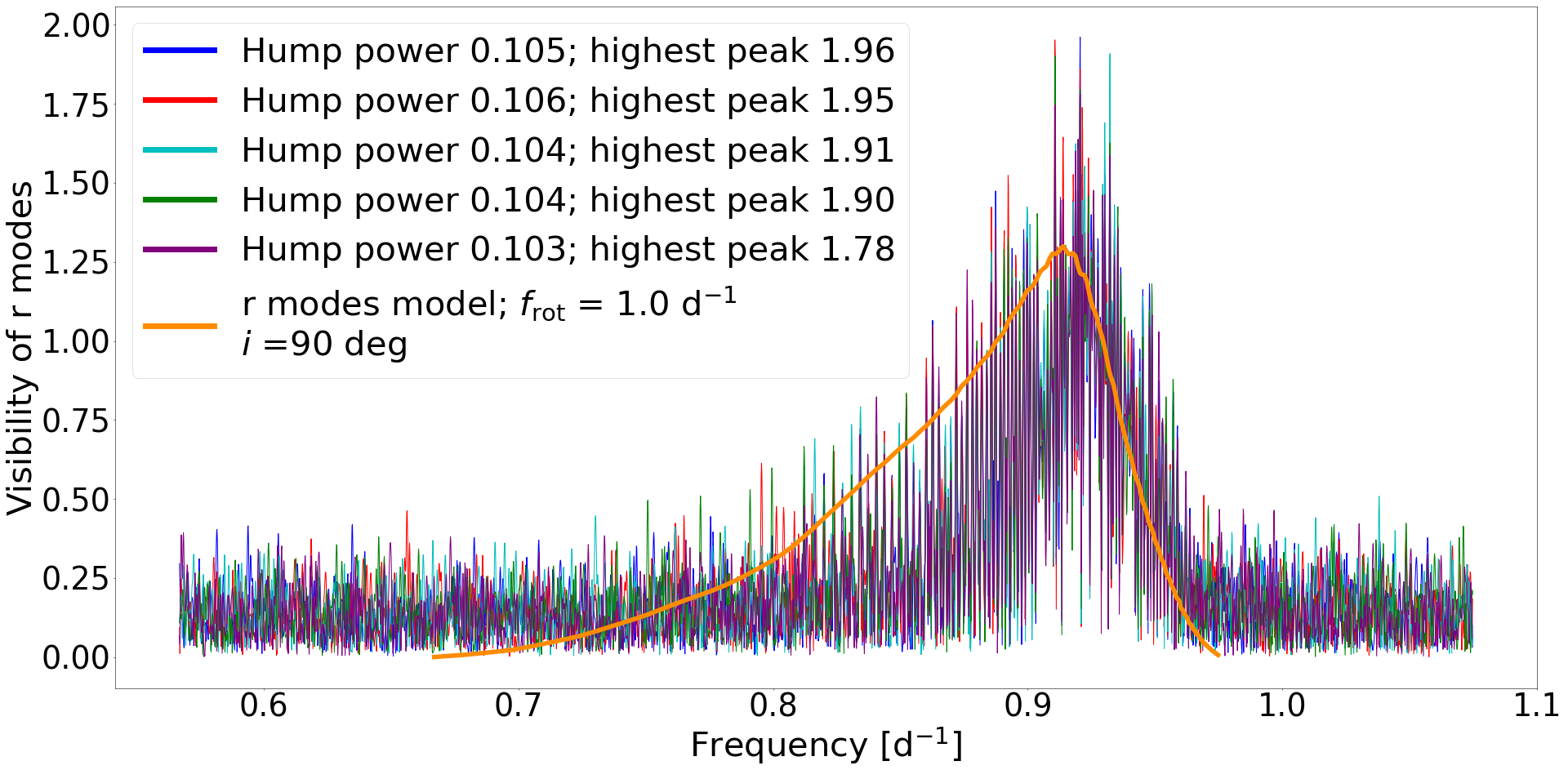}
    \caption{Five realizations of synthetic humps with the same noise level (approx. SNR = 11). The power and highest peak for each realization are displayed in the legend. Due to noise the highest peak is not a representative measure for the size of the hump.}
    \label{fig:hump_area_noise_peak}
\end{figure}

In order to quantify the \textit{hump} features (whether r~ or g~modes), we calculated the power of the humps. We chose to do so because the hump is unresolved and the highest peak is not a representative measure for the hump, as the amplitude of a peak may change due to noise. We illustrate this aspect in Fig. \ref{fig:hump_area_noise_peak}, where for the same noise level but different realizations, the highest peak in the hump varies significantly, while the variation in power only slightly.
To simulate these data, we use the same model ($f_{\rm rot} = 1.0\,{\rm d}^{-1}$) as in section \ref{sec:r_mode_detect}.

The power under the hump was calculated using the integrate.simpson function from \textit{scipy} python package \citep{2020NatMe..17..261V}, which implements the composite Simpson’s rule. This was done directly on the Fourier spectrum. We have accounted for the fact that the noise level in the Fourier spectrum differs from star to star by computing the noise in each observed spectrum and then subtracting its power from the power contained by the hump.
The noise level was estimated in a frequency range between $\approx 15\,{\rm d^{-1}} $ to $\approx 24\,{\rm d^{-1}} $. The Fourier spectrum of each target was inspected independently to ensure that no astrophysical signal was included in the noise determination. We then computed artificial light curves that contained only the noise levels using the \textit{Kepler} time stamps of each star and determined the power (amplitude squared - ${\rm ppm^2}$) contained in the frequency range defined by the humps. Furthermore, to account for the shorter time series in some cases, we have divided the calculated hump power by the power of the spectral window, which we calculated with equation (\ref{eq:psw}).

\begin{equation}
    \label{eq:psw}
    \mathrm{Pw_{SW} = \frac{1}{N \,\Delta\,T},}
\end{equation}
where N is equal to the total number of data points in the time series and $\,\Delta\,T$ is the median value of the cadence of the time series. The resulting unit of the power under the hump is, therefore, $\mathrm{ppm^2/d^{-1}}$.

\subsection{Spectroscopic data}
\label{sec:spectro}

We have analysed spectroscopic data for 28 \textit{hump and spike} stars.
Most spectroscopic data were acquired with the optical echelle spectrograph FIES, located at the 2.56 m Nordic Optical Telescope in La Palma, Spain. The FIES spectra were obtained through dedicated proposals during the observing periods 64 and 65, using the high-resolution fiber (\#4), which offers a spectral resolution of R $\sim 67,000$ \citep{2014AN....335...41T}. The data products used in our analysis were the output of the automated data reduction pipeline - FIEStools (v. 1.5.2)\footnote[2]{\url{www.not.iac.es/instruments/fies/fiestool/FIEStool.html}}. \par
Additional spectra were acquired for a few stars from our sample with the Hertzsprung SONG Telescope (a node of the Stellar Observations Network Group) located at Observatorio del Teide
\citep{2019PASP..131d5003F}. The spectra were obtained with the 1.2 arcsec slit, corresponding to a resolution of R $ \sim 90,000$. \par
We have also made use of archival data obtained with the high-resolution spectrograph HERMES (R $\sim 85,000$) at the Mercator telescope \citep{2011A&A...526A..69R}.  
The meta-data (filename, instrument, date of acquisition, and exposure time) of the spectra analysed in the current work can be consulted in Table \ref{tab:spectra_metadata}. \par
The spectroscopic data served two purposes: 1) to detect any radial velocity shift that might indicate the presence of a stellar companion; 2) to measure the projected rotational velocity, which combined with the rotational period obtained from photometric data from the spike frequency, would help in estimating the stellar inclination. 

In addition to the spectroscopic data analysed in this work, we have also gathered \textit{v}\,sin\,\textit{i} values reported in literature for 61 targets from our sample. The sources of the \textit{v}\,sin\,\textit{i}  values are given in Table \ref{tab:vsini_incl}.

In the following sections, we describe in more depth the spectroscopic analysis methods used to achieve these goals. 

\subsubsection{Synthetic spectra}
\label{sec:synth_spectra}
Synthetic spectra were use as templates in determining radial velocities and aided in the spectral line selection, when estimating the projected rotational velocities (\textit{v}\,sin\,\textit{i}). 
The spectra were computed with the stellar spectral synthesis program SPECTRUM v2.77c \citep{1994AJ....107..742G} \footnote[3]{\url{www.appstate.edu/~grayro/spectrum/spectrum.html}}. When computing the synthetic spectra, ATLAS9  stellar atmosphere models were used, which were downloaded from \footnote[4]{\url{https://wwwuser.oats.inaf.it/castelli/grids/gridp05k2odfnew/ap05k2tab.html}}\citep{2003IAUS..210P.A20C}.

For choosing an appropriate stellar atmosphere model that would be representative for each star, the $T_{\rm eff}$ and ${\rm log} g$ values were used. $T_{\rm eff}$ values were taken from \textit{Gaia} DR2 \citep{2018A&A...616A...1G} and can be found in Table \ref{tab:spike_hump_stellar_param}. The ${\rm log} g$ values can be found in Table \ref{tab:vsini_kics}, where the source is indicated in the footnote. 

The synthetic spectra were computed in the wavelength range from 4000 to 6800~\r{A}. The obtained spectra were also rotationally broadened with the AVSINI function in SPECTRUM with 30, 65, 100, and 150 $\rm km\,s^{-1}$. These values were chosen as the average $v_{\rm rot}$ value for \textit{hump and spike} stars is $\sim\,150\,{\rm km\,s^{-1}}$ (see Table \ref{tab:spike_hump_stellar_param} and figure 8 in \citealt{Henriksen23}), while 100, 65 and 30 ${\rm km\,s^{-1}}$ are representative values for the slower rotating star. The limb darkening coefficient used was 0.6 for all spectra (linear limb-darkening law, see eq. 17.11 in \citealt{2005oasp.book.....G}). This value is a good approximation given the $T_{\rm eff}$ range in which our stars lie (see e.g. \citet{2005oasp.book.....G} p. 437). In the following section, we describe the observed spectroscopic data analysis.

\subsubsection{Radial velocity}
For stars with spectra observed on at least two different nights, radial velocity (RV) shifts were determined. The RV shifts between the two epochs were extracted by cross-correlating the observed with corresponding synthetic spectra, using the function \textit{cross\_correlate\_with\_template}, from the iSpec software solution \citep{2014A&A...569A.111B,2019MNRAS.486.2075B}. For the FIES and SONG spectra, the cross correlation was performed for each echelle order. The final RV value and its uncertainty were calculated as an weighted average of the individual RVs from each echelle order and as the error on the weighted mean, respectively, using equations (\ref{eq:rv}) and (\ref{eq:rv_err}). The RV uncertainties for each échelle order ($\sigma_{\rm RV_{order}}$) are calculated through iSpec, which implements the maximum-likelihood approach given by \citet{2003MNRAS.342.1291Z}. The fitting model used in this work was a $2^{\rm nd}$ order polynomial and a Gaussian. The reader is referred to the iSpec documentation for more details \citep{2014A&A...569A.111B,2019MNRAS.486.2075B}.

\begin{equation}
    \label{eq:rv}
    \rm{RV_{final}} = \frac{\sum RV_{order}/ \sigma_{RV_{order}}^{-2}}{\sum \sigma_{RV_{order}}^{-2}} 
\end{equation}

\begin{equation}
    \label{eq:rv_err}
    \sigma_{\rm{RV_{final}}} = \frac{1}{\sum\sigma_{\rm RV_{order}}^{-2}} 
\end{equation}

Orders that contained telluric lines or broad lines (such as: H$\alpha$ or H$\beta$), or which did not contain any spectral lines, were excluded. The telluric lines atlas that was used for guidance is the telluric mask which iSpec provides \citep{2014A&A...569A.111B} and which is based on synthetic spectrum described in more detail in \citet{2014AA...564A..46B}.
 
In the case of HERMES spectra, we worked with order-merged, 1D wavelength calibrated spectra. For computation purposes, and in order to avoid regions where telluric lines were present, the spectra were clipped into 13 sections (Table \ref{tab:hermes_sections} contains the wavelength limits of these sections). The final RV values were computed as the weighted average from those obtained from each of the 13 sections. The results of our analysis can be seen in Table \ref{tab:rv_results_multiple_epochs}, for stars with two or more available epochs, and in Table \ref{tab:rv_results_one_epoch}, where only one epoch was observed. We consider epochs separate when the data originate from different observing nights. The instrument used is specified in the aforementioned tables. The results in Table \ref{tab:rv_results_multiple_epochs} indicate that out of the 22 stars for which a radial velocity shifts could be determined, 11 stars show signs for possible stellar companions. We classified a star as binary when the RV measurements from the different epochs were discrepant based on the $3\sigma$ uncertainties. We note that the stars indicated as non-binary, could still have a companion that eluded our observations.

This means that 11 stars out of 22 \textit{hump and spike} stars for which we could gather RV information, could be part of binary (multiple) systems. Depending on the literature source the binary occurrence in chemically normal A stars is between 35 and 60 per cent (\citealt{2009AJ....138...28A, 2008MNRAS.389..869E,2002AJ....123.1570P}) which is consistent with our results.  
However, we consider that a more extensive study regarding the binary occurrence is needed, as the number of stars is too small and additional data points are needed to establish the presence of stellar companions. 

The results presented in Table \ref{tab:rv_results_one_epoch} were used to shift the observed spectra to align  with the synthetic spectra, for an easier identification of spectral lines that could be used in computed the projected rotational velocities. This method is described in the following subsection.

\subsubsection{\textit{v}\,sin\,\textit{i} determination - the Fourier method}
\begin{figure}

	\includegraphics[width=\columnwidth]{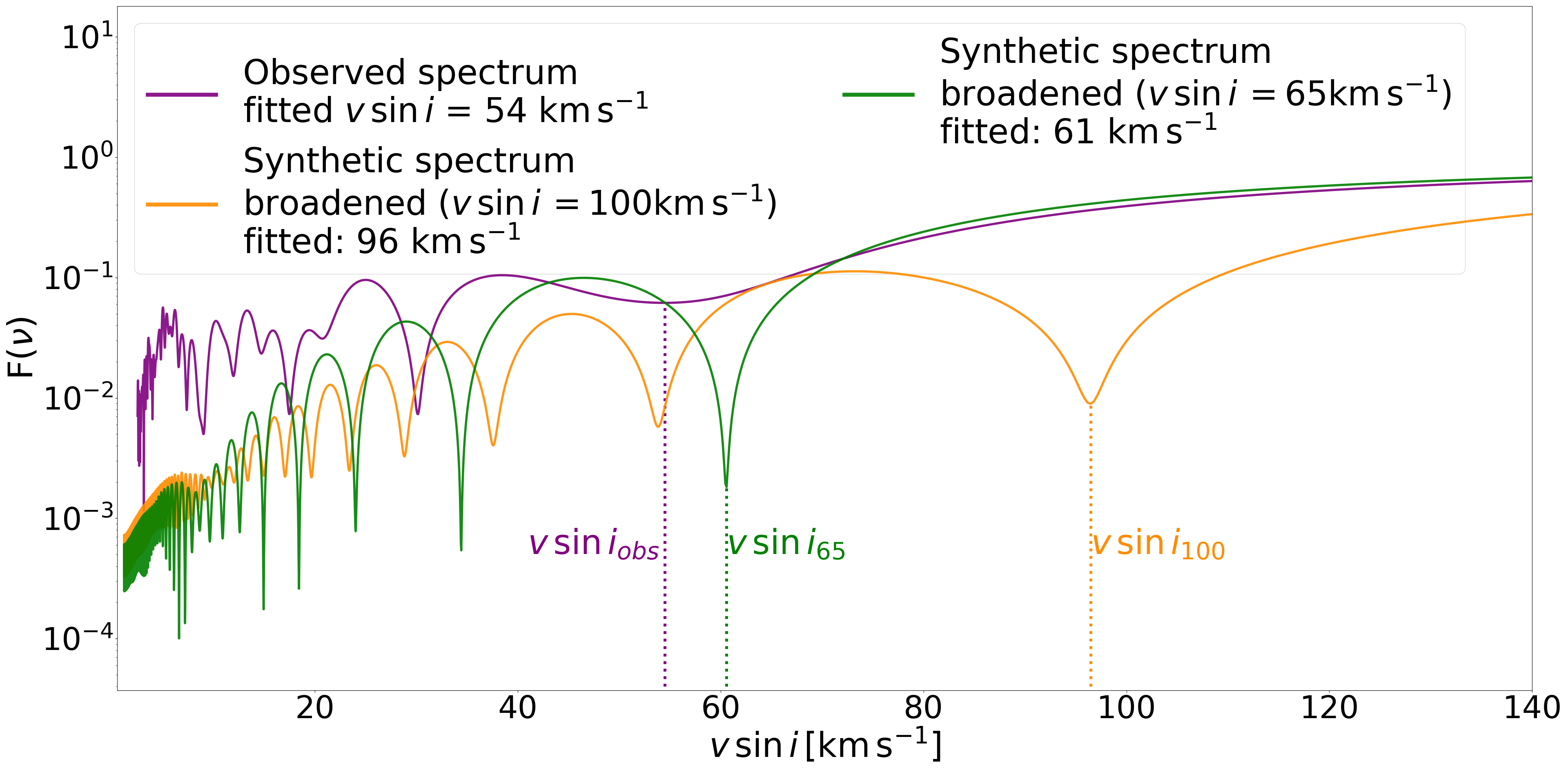}
 	\includegraphics[width=\columnwidth]{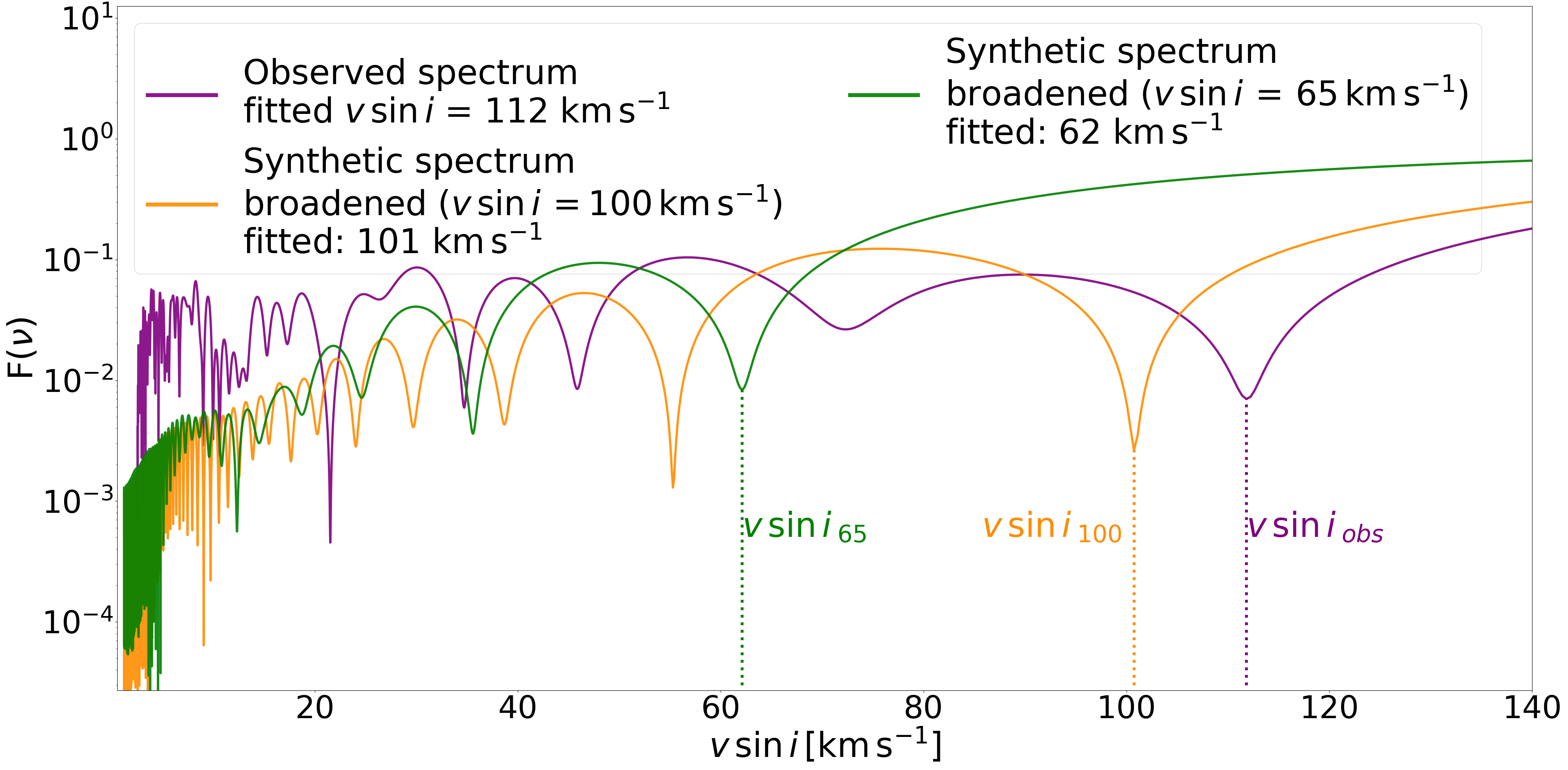}
    \caption{Fourier transform of the rotation profile for the spectral line Fe I 6065.48~\r{A}. 
    \textit{Upper panel}: KIC\,7116117 - Fourier transform for the observed spectral line and the spectral line of two synthetic spectra that were rotationally broadened with $65{\rm\,km\ s^{-1}}$ and $100{\rm\,km\ s^{-1}}$. \textit{Lower panel}: KIC\,5938266 - Fourier transform for the observed spectral line and the spectral line of two synthetic spectra that were rotationally broadened with $65{\rm\,km\ s^{-1}}$ and $100{\rm\,km\ s^{-1}}$. We note that both panels are similar to figure 3 from \citep{2002A&A...381..105R}, however, we prefer to display the inverse of the frequency obtained through the Fourier transform, as it is equivalent to the \textit{v}\,sin\,\textit{i} value.}
    \label{fig:vsini_fourier_curves}
\end{figure}

To extract the projected rotational velocities (\textit{v}\,sin\,\textit{i}) from the available spectroscopic data, the Fourier technique was adopted.  As described by \citet{1976PASP...88..809S} and applied in works such as \citet{2002A&A...381..105R, 2002A&A...393..897R} or \citet{2006A&A...448..351S}, the Fourier transform of a spectral line produces a curve with sidelobes that alternate in being either dips or peaks and will gradually flatten in amplitude. Examples of such curves are displayed in Fig. \ref{fig:vsini_fourier_curves}, where the Fourier transforms of an iron line (Fe I 6065.48~\r{A}), found in the observed spectra of KIC\,7116117 and KIC\,5938266, are compared with the Fourier transforms of the corresponding synthetic spectral line found in rotationally broadened synthetic spectra. The frequency of the first zero (dip) in the Fourier transform of the line profile corresponds to the last dip in velocity space which indicates the value of the \textit{v}\,sin\,\textit{i}. The Fourier transform was done using the spectra.tools.vsini function from the Institute of Astronomy at KU Leuven (IvS) python package \footnote[5]{\url{https://github.com/IvS-KULeuven/IvSPythonRepository}}. If multiple spectra were available for a target, the technique was applied on the stacked spectrum. The spectra were combined by using the weighted average flux values, if flux errors were available (FIES and SONG spectra) or only the average flux values (HERMES spectra). 

We tested the Fourier technique on two stars with known \textit{v}\,sin\,\textit{i} values from literature with which we compared our results. KIC\,9117875, a relatively slow rotator (with respect to the most stars in our sample), has a literature \textit{v}sini \textit{i} value of $61 \pm 3\,{\rm km\ s^{-1}}$ \citep{2015MNRAS.450.2764N}.We find an average of $57\pm 6\,{\rm km\ s^{-1}}$, based on four lines: Fe I 5633.98~\r{A} ($65.5\,{\rm km\ s^{-1}}$), Fe I 6065.48~\r{A} ($54.2\,{\rm km\ s^{-1}}$), Fe I 6078.49~\r{A} ($58.1\,{\rm km\ s^{-1}}$), Fe I 6393.61~\r{A} ($50.5 \,{\rm km\ s^{-1}}$). The second test star to validate the method, is KIC\,4572373, a fast rotator, and has been reported to have a projected rotational velocity value of $184 \pm 7$ \citep{2015MNRAS.450.2764N}. Given its $T_{\rm eff}$ and fast rotation, there were not many lines to choose from. We used Mn II 4481.13/33 ($173\,{\rm km\ s^{-1}}$) and Si II 6371.35 ($215{\rm\,km\ s^{-1}}$) to derive an average \textit{v}sini \textit{i} value of $194 \pm 21\,{\rm km\ s^{-1}}$, which is in agreement with the value reported in literature. This indicates that our method of measuring \textit{v}sini \textit{i} from spectral lines yields reliable results. 

\begin{figure}

	\includegraphics[width=\columnwidth]{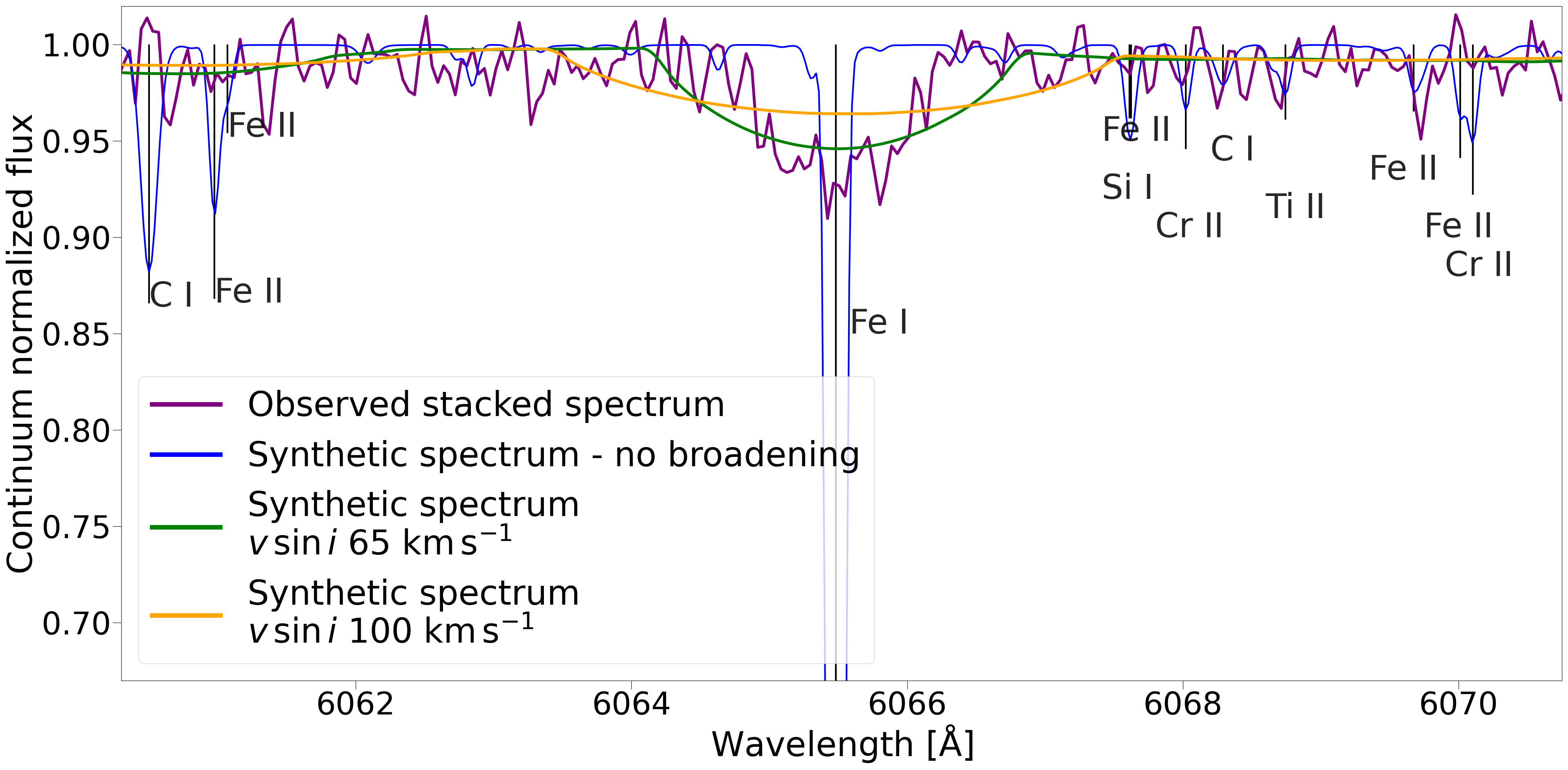}
 	\includegraphics[width=\columnwidth]{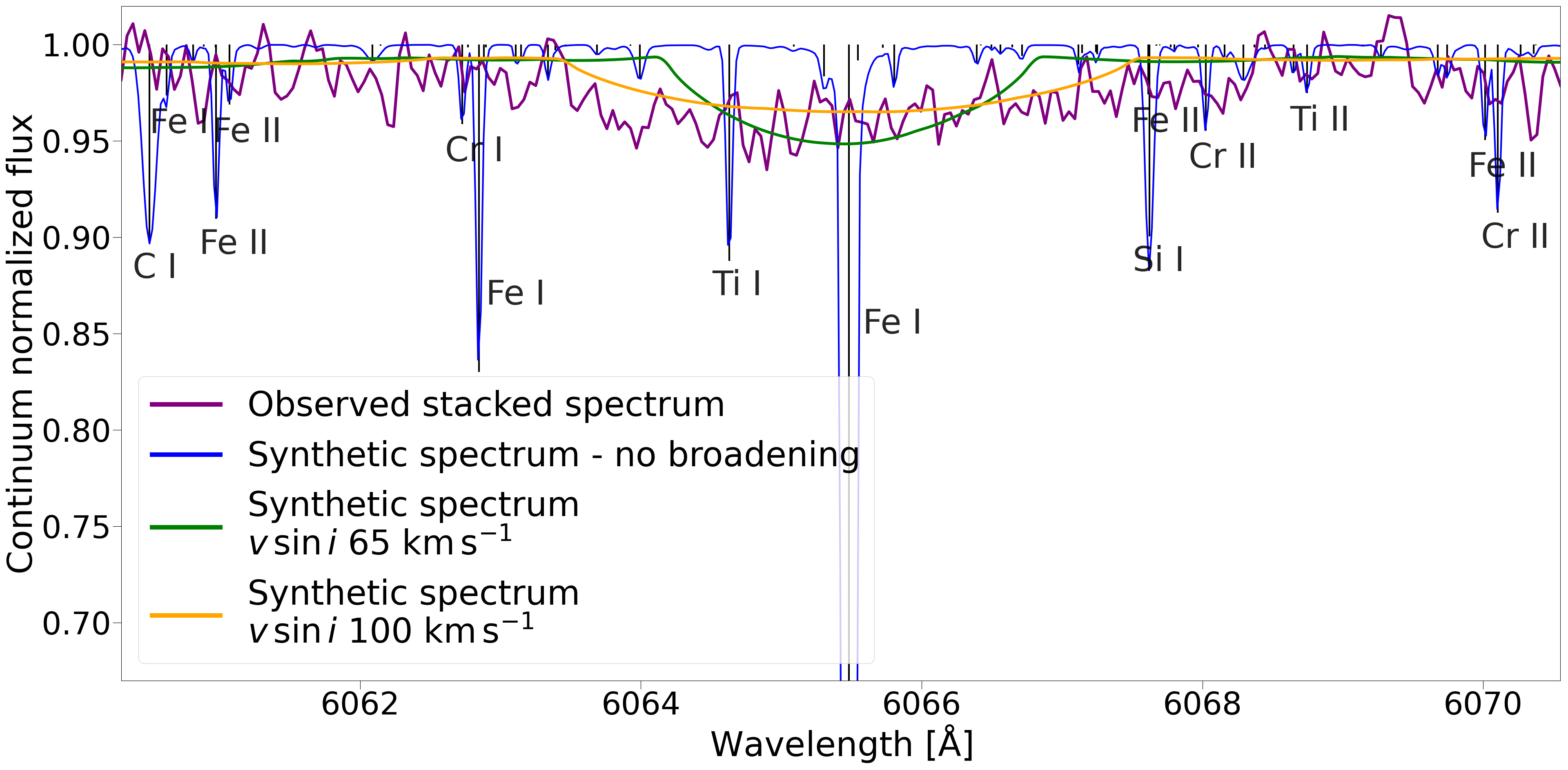}
    \caption{Observed and synthetic spectral line -  Fe I 6065.48 \r{A}. Synthetic lines rotationally broadened by $65{\rm\,km\ s^{-1}}$ and $100{\rm\,km\ s^{-1}}$ are also displayed for comparison. For clarity only strong lines are labeled. Both spectra are taken with FIES.
    \textit{Upper panel}: KIC\,7116117. \textit{Lower panel}: KIC\,5938266. One can see the appearance of more species as this target is$\sim 1200$\,K cooler than the one in the upper panel. }
    \label{fig:line6065_example}
\end{figure}

The selection of the wavelength range that would define a spectral line was done manually for each star and line. This was due to the fact that the wavelength range that described a spectral line depended on the rotational broadening of the star. In Fig. \ref{fig:line6065_example}, the Fe I 6065.48 \r{A} spectra line is displayed for two of our analysed targets:  KIC 7116117 and KIC 5938266. A synthetic spectrum is also depicted together with the observed data. The synthetic spectrum was rotationally broadened with indicated values as described in section \ref{sec:synth_spectra}. The species displayed were obtained with the LINES function from the SPECTRUM software solution. The atomic data file used in compiling a line list,  was the standard atomic and molecular data file, provided by SPECTRUM and which is based on \citealt{1998SSRv...85..161G}. The vertical gray lines show the contribution of the species indicated by the annotated texts. The location and the nature of the spectral lines is useful when choosing adequate spectral lines for determining the rotational broadening, as lines that are fairly strong and "isolated" (less contaminated) as possible are preferred. This, however, is not always possible, as in some cases the high rotation makes it impossible to find "ideal" lines, as they are either not very strong or contaminated. This task has proved to be difficult for both hot stars, where less species can be found in comparison with cooler counterparts, and for cool stars, where presence of more spectral lines implies higher chance of blending (e.g. see the difference in the same wavelength region of the observed spectrum of KIC\,7116117 and KIC\,5938266 in Fig. \ref{fig:line6065_example}). 

The results of our \textit{v}\,sin\,\textit{i} analysis using the Fourier technique can be found in Table \ref{tab:vsini_kics}, where the average \textit{v}\,sin\,\textit{i} values and the number of lines used are listed for each star. In the Appendix, the \textit{v}\,sin\,\textit{i} computed from each line can be found (Table \ref{tab:vsini_lines_kic_1}). Note that uncertainties reported in Table \ref{tab:vsini_kics} are the standard deviations of the mean values. For this reason, the uncertainties are underestimated for some targets for which \textit{v}\,sin\,\textit{i} measurements are accurate.

For KIC\,5903499 and KIC\,6974705 we note that the line selection was restricted by the high rotation rate and $T_{\rm eff}$ of the stars (see Table \ref{tab:spike_hump_stellar_param}). Using the rotational frequency from the spike and the stellar radii from \citet{2020AJ....159..280B}, we derive rotational velocities of $200^{+10}_{-9} {\rm\,km \ s^{-1}}$ and $196^{+7}_{-6}{\rm\,km \ s^{-1}}$, respectively. The \textit{v}\,sin\,\textit{i} values could be measured based on only one line (the doublet Mn 4481.13/33). Furthermore these two targets are relatively faint (KIC\,5903499 - V=10.7; KIC\,6974705 - V = 10.5). Both targets were planned to be observed with FIES in two separate epochs in our observing proposal, but one of the planned observations was lost due to bad weather. This means that the observed spectra have also low SNR and measuring the rotational broadening on spectral lines is difficult. In fact for KIC\,5903499 it was very hard to measure the radial velocity shift with respect to the synthetic spectra, due to the lack of resolved spectral lines. Only three of the échelle orders were used.  

We have a similar remark regarding KIC\,8037519. The combination with high $T_{\rm eff}$ and $v_{\rm rot}$ (Table \ref{tab:spike_hump_stellar_param}), made it difficult to find spectral lines. We give an estimate \textit{v}\,sin\,\textit{i} based on the Mn $4481.13/33$~\r{A}  doublet line and the $6371.35$~\r{A} Si II line. We note that the Si line has some contamination from other species, but given the fast rotation of this star, the blending effect is negligible. The obtained values can be consulted in Table \ref{tab:vsini_lines_kic_1}.

We also note that for KIC\,5024410 and KIC\,7939835, for which we measure low projected rotational velocities, could have some problems. As noted in \citealt{2005oasp.book.....G}, for slow rotators, other broadening factors such as micro- or macro-turbulence become significant and should be accounted for when fitting the rotational broadening of spectral lines. In other words, our results for these relatively slow rotators (relative with respect to the other \textit{hump and spike} stars for which the rotational velocity spans from $\sim 50 {\rm \,km\ s^{-1}}$ to $\sim 300 \,\rm{km\ s^{-1}}$, see fig. 8 in \citealt{Henriksen23}), might not be as accurate and should be regarded with caution. More in-depth analysis should be conducted for these targets, however, we consider these values to be sufficiently accurate for the scope of this work. 

\begingroup
\renewcommand{\arraystretch}{0.879}
\begin{table}

	\caption{Overview of RV results for stars with multiple epochs. The instrument with which the data were obtained is indicated in parenthesis under the KIC ID. *more data available for this target. Table \ref{tab:spectra_metadata} contains the list of all available spectra. }

	\label{tab:rv_results_multiple_epochs}
	\begin{tabular}{|cccc|c|} 
		\hline
    KIC& Midpoint of obs.&  RV  & 3 $\sigma_{\rm RV}$  &  Binary?\\
    
    (Instrument)& [Y-M-D h:m] UT& [ $\rm km\,s^{-1}$] &[ $\rm km\,s^{-1}$] & \\
    \hline
    
    3634487& 2022-06-20 00:36&6&8.6& n \\
    (FIES)&2022-05-27 00:38&13&7.2&  \\
   \hline
    3848948 & 2022-06-20 03:55 & -6& 5.7& n\\
     
    (FIES) & 2022-05-26 02:10 &0 & 1.6& \\
    \hline
    4572373 & 2012-07-14 01:55 & 9& 7.3& n \\
     
    (HERMES)&  2012-09-05 00:43 & 4& 8.4& \\
    &  2012-09-05 01:13 & 0& 2.9& \\
    \hline
    4661914 & 2022-06-20 02:14& -21 & 1.6 & y \\ 
     
    (FIES) & 2022-05-26 04:06 &-10 & 1.4 & \\ 
    \hline
    4856799 & 2022-06-20 04:53 &-17 & 9.8  &y\\
     
    (FIES) & 2022-05-27 05:08 & -2&  3.9&\\

    \hline 

   5024410& 2013-07-26 22:47 & 3 & 0.03 & y\\

   (HERMES)*& 2013-07-28 01:28 & 4 & 0.03 & \\

   & 2013-07-30 00:14 & 2 & 0.03 & \\
   & 2013-10-01 20:55 & -9 & 0.03 & \\

   & 2013-10-04 21:42 & -10 & 0.03 & \\
   & 2021-08-21 02:11 & -2 & 0.03 & \\
    \hline

    5456027&2022-06-20 01:50&-2 &  2.8 & n\\
     
    (FIES)&2022-05-26 04:52&  -0&  1.5 & \\

    \hline
    5524045 & 2011-09-19 22:37& -4 & 4.5 & n\\
     
    (HERMES)&  2018-11-06 22:05 & -5 & 3.3 & \\
    &  2018-11-06 22:26 & -3 &  3.1 & \\

    \hline
    5980337& 2022-06-20 02:37 & 10 & 3.0& y\\
     
    (FIES)& 2022-05-26 02:36 & 22  & 3.0 & \\
    \hline 
    6039039& 2022-06-21 01:19 & 0& 2.3 & y \\
    
    (FIES)& 2022-05-27 03:41 & 11&2.5 & \\

    \hline 
    6610433& 2022-06-06 04:35 & -17&  4.3 & n\\
     
    (FIES)& 2022-06-21 00:49 & -9&  4.6 & \\

    \hline
    7116117& 2022-06-20 01:04 &-21& 0.8 & y\\
     
    (FIES)& 2022-05-27 04:40 & -10 & 0.7 &\\
    \hline 
    7131587 & 2021-10-04 21:09 & 30 & 1.1 & n\\ 
     
    (FIES) & 2021-10-10 21:54 & 29 & 1.1 &\\
    \hline 
    7842339 & 2021-10-04 21:42 &10 & 4.4  & n\\
     
    (FIES) & 2021-10-10 22:32 &17 & 5.3  & \\
    
    \hline
    7939835& 2022-06-21 00:22 & -5& 0.3 & y\\
     
    (FIES) &2022-05-27 00:11 & 7& 0.3 & \\
    
    \hline
    7959579 & 2022-06-20 01:26  &-14 & 4.9  & y\\
     
    (FIES) & 2022-05-26 04:27  & -1& 1.4  &\\
        \hline
    8037519 &2018-09-06 21:32 & 15 & 12.0 & y \\
     
    (SONG) & 2021-06-16 01:20 & -11 & 9.6 & \\
    \cline{2-4}
    \multirow{2}{4em}{(HERMES)} & 2016-04-12 05:08 & 30 & 19.8&\\
    & 2016-04-13 03:37 &29& 18.7 &\\
    \hline
    9117875& 2011-07-11 22:30& 0 & 1.0 & n \\
    (HERMES)& 2011-07-12 22:33& 0 & 0.9 & \\
    \hline 
    9273647 & 2022-06-20 03:34  &-26& 2.4& y\\
     
    (FIES) & 2022-05-26 01:48  &-15 & 2.4   &   \\
    \hline 

    9519698 & 2022-06-19 05:15 &-3 &  7.2  & n\\
     
    (FIES) & 2022-05-26 03:29  &  1 & 4.6   &\\

    \hline
    10068389 &2022-06-19 05:01 & -34 & 7.0& n\\
     
    (FIES) &2022-05-26 03:44 & -33& 7.3 & \\

    \hline
    10810140 & 2022-06-20 04:24& -11 & 5.0 & y\\
     
    (FIES) & 2022-05-27 04:12& -1 & 2.1& \\

    \hline

    \end{tabular}
\end{table}
\endgroup

\begingroup
\renewcommand{\arraystretch}{0.85}
\begin{table}
\centering
	\caption{Overview of RV results for stars with one epoch. The data are sorted by instrument. For targets with multiple observations the listed RV and uncertainties values are average values.}

	\label{tab:rv_results_one_epoch}
	\begin{tabular}{cccc} 
		\hline
    KIC& Midpoint of obs.&  RV  & 3 $\sigma_{\rm RV}$ \\
    & [Y-M-D h:m] UT& [ $\rm km\,s^{-1}$] &[ $\rm km\,s^{-1}$] \\
    \hline
    & FIES &  & \\
    \hline
    4066110& 2021-10-16 21:29 & 45&1.2 \\
    & 2021-10-16 21:55&& \\
    & 2021-10-16 22:20&& \\
    \hline
    5566579 & 2021-10-03 22:40 & -2 & 2.7 \\
    
    & 2021-10-03 23:09 & &  \\

    \hline
    5903499& 2021-10-14 21:17&0 & 5.3 \\
        &2021-10-14 21:48 & & \\
    \hline
    5938266 & 2021-10-03 21:34 &23 & 1.9\\
     
    & 2021-10-03 22:05 & &  \\
    
    \hline 
    6974705 &2021-10-12 22:24& 11 &  7.6  \\
   
     &2021-10-12 22:50& &   \\  
     \hline 
    8846809& 2021-10-12 21:50 & 1& 1.9 \\
    \hline
    & SONG& &  \\
    \hline
   8783760 & 2021-06-21 01:26 & -6& 6.3 \\
   \hline
   9163520 & 2018-09-28 23:27 & -19 &  0.4\\
           & 2018-09-28 23:35 & & \\
           & 2018-09-28 23:43 & & \\
    \hline
    & HERMES& &  \\
    \hline
    6192566 & 2011-09-19 21:01 & -29 & 2.4  \\ 

   \hline
    \end{tabular}
\end{table}
\endgroup

\begingroup
\renewcommand{\arraystretch}{0.95}
\begin{table}
\centering
\caption{Extracted \textit{v}\,sin\,\textit{i} values with Fourier technique. The superscript in the ${\rm log} g$ column indicates the source of the values. The last column indicates the number of lines on which the Fourier transform was applied and contributed to obtaining the final \textit{v}\,sin\,\textit{i} values.}
\label{tab:vsini_kics}
\begin{tabular}{|ccccc|} 
\hline
KIC& ${\rm log} g^*$&\textit{v}\,sin\,\textit{i}  & No. \\
& [dex]&[ $\rm km\,s^{-1}$] & lines \\
\hline

3634487& $3.9 \pm 0.08^{\rm m}$&$175 \pm 14$ &4\\
3848948& $4.1 \pm 0.09^{\rm m}$&$98 \pm 12$ &4\\
4066110& $3.5 \pm 0.07^{\rm m}$&$138 \pm 19$ &3\\
4661914& $4.2 \pm 0.1^{\rm m}$&$85 \pm 14$ &4\\
4856799& $4.0 \pm 0.07^{\rm m}$&$173 \pm 14$ &5\\
5024410&$3.8 \pm 0.21^{\rm b}$&$7 \pm 2$ &25\\
5456027& $4.0 \pm 0.09^{\rm m}$&$160 \pm 6$ &4\\
5566579& $4.1 \pm 0.09^{\rm m}$&$141 \pm 20$ &3\\
5903499& $4.2 \pm 0.1^{\rm m}$&$254$  &1\\
5938266& $3.8 \pm 0.08^{\rm m}$&$103 \pm 16$ &4\\
5980337& $3.9\pm0.12^{\rm f}$ &$86 \pm 17$ &4\\
6039039& $3.9 \pm 0.07^{\rm m}$&$77 \pm 6$ &5\\
6192566& $3.6 \pm 0.09^{\rm m}$&$51 \pm 11$ &6\\
6610433& $4.1 \pm 0.08^{\rm m}$&$118 \pm 6$ & 4 \\
6974705& $4.0 \pm 0.08^{\rm m}$&$246 $ & 1 \\
7116117& $4.0 \pm 0.1^{\rm m}$&$70 \pm 10$ & 4 \\
7131587& $4.0 \pm 0.09^{\rm m}$&$49 \pm 8$ & 15 \\
7842339& $3.9 \pm 0.09^{\rm m}$&$104 \pm 11$ & 3\\
7939835& $4.0 \pm 0.09^{\rm m}$&$24 \pm 6$ & 14 \\
7959579& $4.2 \pm 0.1^{\rm m}$&$117 \pm 5$ & 4 \\
8037519&$3.9 \pm 0.05^{\rm b}$&$186 \pm 10$ & 2 \\
8783760& $3.9 \pm 0.09^{\rm m}$&$127 \pm 16$ & 5\\
8846809& $3.7 \pm 0.09^{\rm m}$&$154 \pm 10$ & 3 \\
9163520& $3.5 \pm 0.29^{\rm l} $ & $41\pm4$ & 6 \\
9273647& $4.0 \pm 0.1^{\rm m}$&$116 \pm 6$ & 4 \\
9519698& $4.0 \pm 0.08^{\rm m}$&$175 \pm 6$ & 5 \\
10068389& $3.9 \pm 0.08^{\rm m}$&$132 \pm 9$ & 5 \\
10810140& $4.2 \pm 0.07^{\rm m}$&$130 \pm 5$ & 5 \\

\hline
\multicolumn{4}{l}{* m = \citealt{2019MNRAS.485.2380M}; b = \citealt{2020AJ....159..280B};}  \\
\multicolumn{4}{l}{f = \citealt{2016AA...594A..39F}; l = \citealt{2014AJ....147..137L}} \\

\hline
\end{tabular}
\end{table}
\endgroup

\subsection{Stellar inclination estimate}
\label{sec:incl}

\begin{figure}

	\includegraphics[width=\columnwidth]{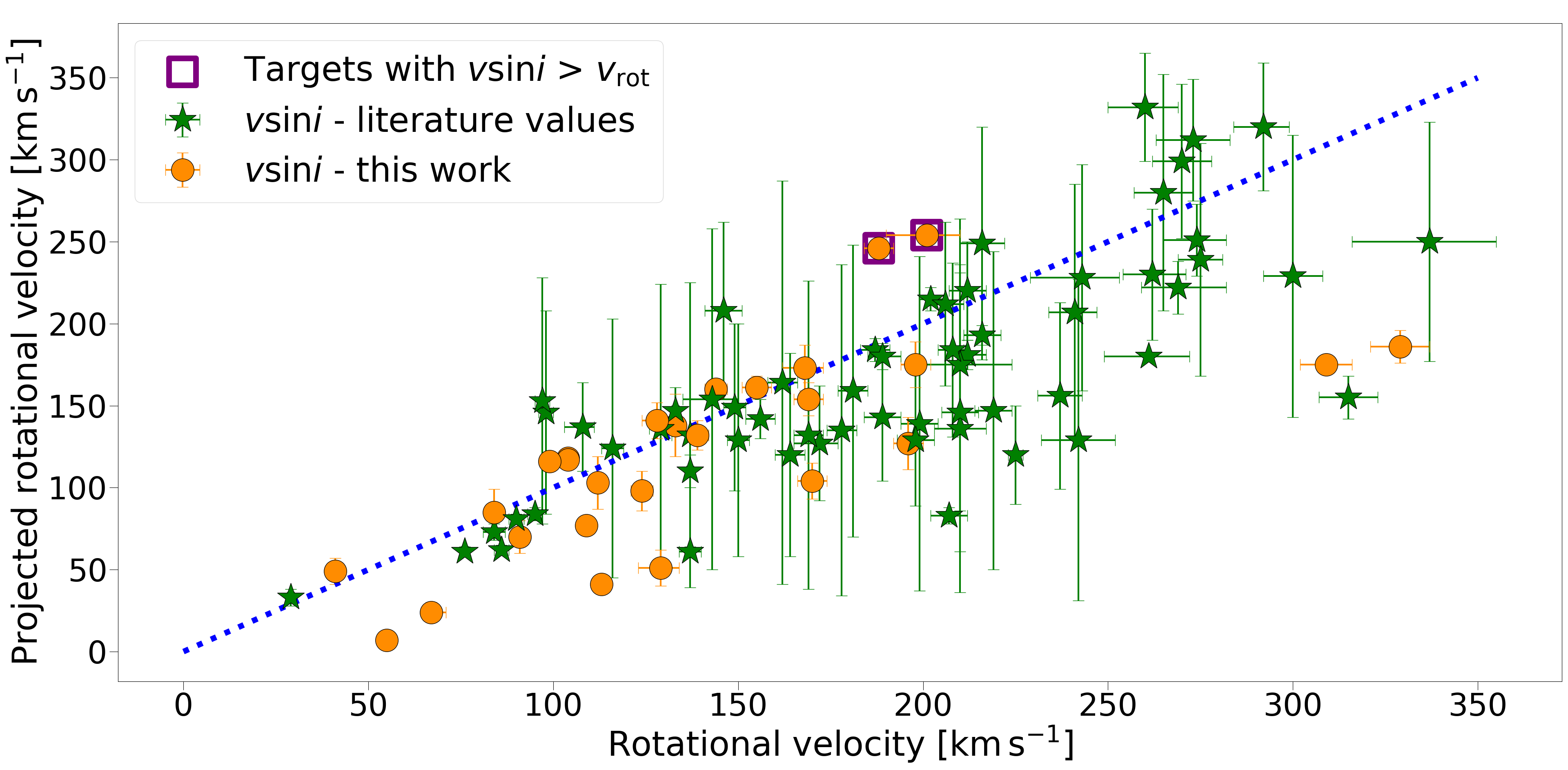}
    \caption{Projected rotational velocity values as a function of rotational velocity values. The dotted blue line stands for the one-to-one relation. Stars for which \textit{v}\,sini\,\textit{i} values where determined in this work are depicted with orange circles. A green star marker indicates targets for which the \textit{v}\,sini\,\textit{i} values is found from literature as indicated in Table \ref{tab:vsini_incl}. Targets where \textit{v}\,sin\,\textit{i} > $v_{\rm rot}$ are highlighted with purple squares. }
    \label{fig:vrot_vs_vsini}
\end{figure}

We show the \textit{v}\,sin\,\textit{i} values (determined here and from literature) as a function of $v_{\rm rot}$ in Fig. \ref{fig:vrot_vs_vsini}. The uncertainties associated with $v_{\rm rot}$ were obtained through error propagation using the radii and $f_{\rm rot}$ uncertainties listed in Table \ref{tab:spike_hump_stellar_param}. There are two stars for which the rotational velocities obtained  do not agree with the measured \textit{v}\,sin\,\textit{i} (highlighted with magenta squares: KIC\,5903499 and KIC\,6974705). 

There are a few reasons that can explain the discrepancy between the $v_{\rm rot}$ and \textit{v}\,sin\,\textit{i} values. Firstly, the uncertainties in the stellar radii, which propagate into the $v_{\rm rot}$ determination, could be underestimated. Secondly, since we could not rule out the presence of stellar companions (both these stars were observed with FIES in one only night, see Table \ref{tab:rv_results_one_epoch}), it is possible for their radii to be inaccurate as they could be binaries. 
Thirdly, as mentioned in the previous section the combination of low SNR, fast rotation and high $T_{\rm eff}$ made it difficult to find spectral lines suitable for the Fourier method. Since we could only use the Mn doublet, we could not compute a standard deviation as an estimate for the uncertainty. As \citet{2002A&A...381..105R} noted, the \textit{v}\,sin\,\textit{i} error is expected to increase as the nominal value increases, because spectral lines become shallower and the occurrence of line blending increases. We choose to set the inclination values for these two stars to 90 deg. In Fig. \ref{fig:area_hump_incl}, they are highlighted with magenta circles and are not taken into account when the correlation factor is calculated.

For the remaining stars, we have calculated the stellar inclination (\textit{i}) with the equation:
\begin{equation}
    i = {\rm arcsin } \left(\frac{v\,{\rm sin}\,i}{v_{\rm rot}}\right)
\end{equation}

Given the non-linearity of the arcsin function and its asymptotic behaviour for values close to unity, calculating the uncertainties associated with the stellar inclination cannot be done by error propagation assuming a linear combination of variables, or by approximating with a first-order Taylor series expansion. Furthermore, in the case of some stars, the \textit{v}\,sin\,\textit{i} values are higher than the $v_{\rm rot}$ values, which makes it impossible to calculate the inclination, as arcsin is not defined for values higher than unity. For this reason we have chosen to estimate the stellar inclination and its associated uncertainty by random sampling, drawing $10^7$ values for both \textit{v}\,sin\,\textit{i} and $v_{\rm rot}$, from truncated normal distributions centered at the nominal values and with width of the associated uncertainties ($3\,\sigma$). The reported values in Table \ref{tab:vsini_incl} are the mean and standard deviation of the resulting distribution of calculated inclinations.

We note that despite measuring a \textit{v}\,sini\,\textit{i} value for KIC\,5980337, we do not report an inclination value as it cannot be assessed which of the two spikes should be used in calculating the $v_{\rm rot}$ (see Fig.\,\ref{fig:example_hump_spike}).

\section{Results and discussion}
\label{sec:result}
In this section we present our results based on photometric and spectroscopic observations. From photometric data we have extracted the power of the r and g~modes humps and we used the spike amplitudes and frequencies as derived in \citet{Henriksen23}. Combining spectroscopic and photometric data has allowed us to compute estimates of stellar inclination values. We show various correlations based on parameters described in the previous sections, as well as other stellar parameters, such as $T_{\rm eff}$ or luminosity, using similar sources as in \citet{Henriksen23}.

\begin{figure}

	\includegraphics[width=\columnwidth]{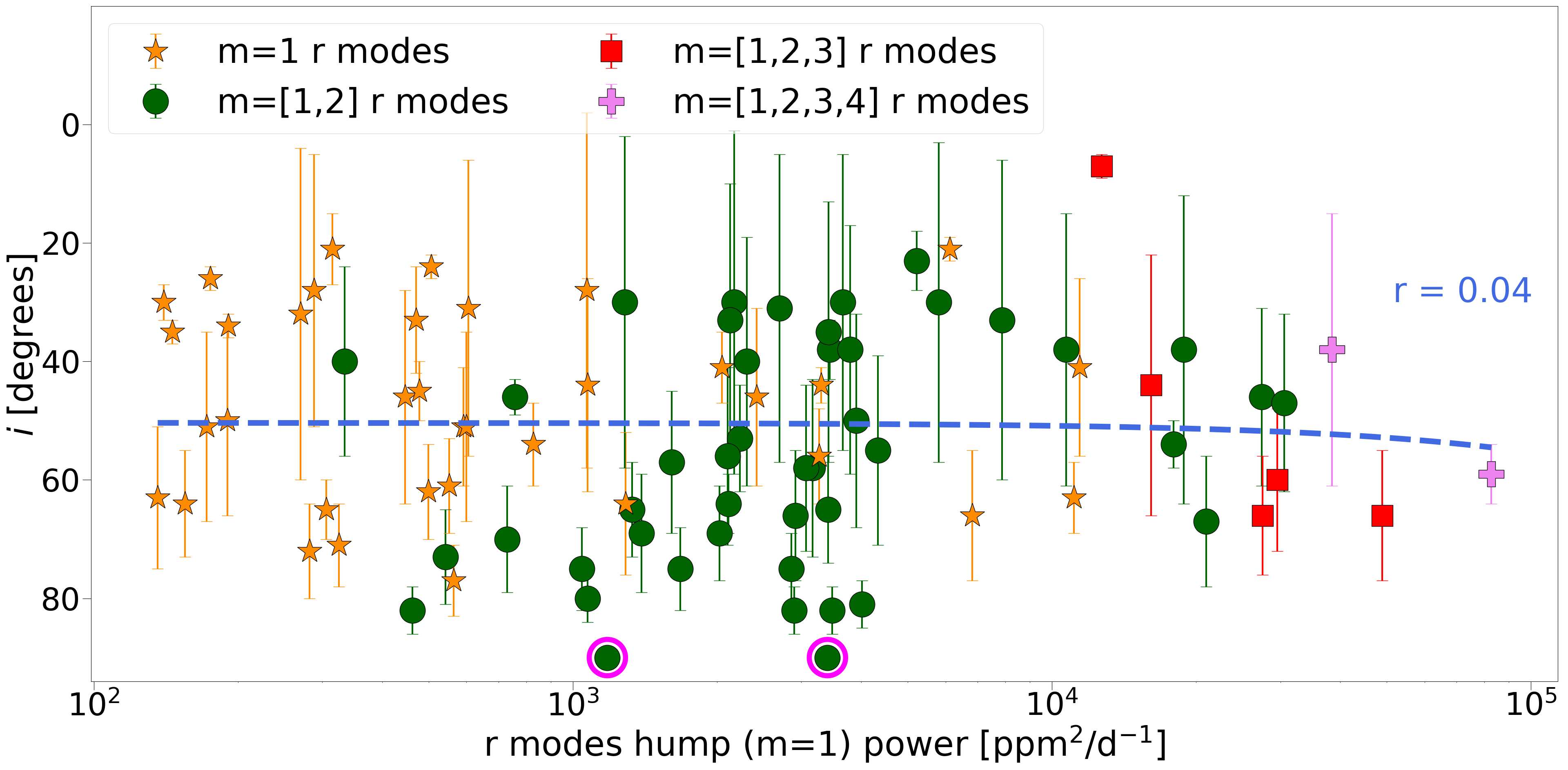}
    \caption{The power in the main hump as a function of the stellar inclination. Targets circled with magenta have the inclination set to 90 degrees. See section \ref{sec:incl} for more details. The symbol type denotes how many r~mode humps a star possesses (e.g., the orange star symbol indicates that a target exhibits only one $m=1$ r~mode hump, while green circles indicate that a star exhibits both $m=1$ and $m=2$ r~mode humps.) }
    \label{fig:area_hump_incl}
\end{figure}

In Fig. \ref{fig:area_hump_incl} we depict the stellar inclination as a function of the power described by the main hump ($m=1$ r modes). The colour-code of the figure suggests that the inclination does not play a role in the visibility of the number of azimuthal orders excited in the star. The notable trend is that, with the exception of one star, all star have an inclination $>20\,\rm{deg}$. Furthermore, it seems that stars with very large power in their ($m=1$) r~mode humps have intermediate inclination values (approx. $55\pm10$ deg). This is in agreement with r~modes tending to be confined in mid-latitudinal regions rather than the equator (\citealt{1997ApJ...491..839L,2018MNRAS.474.2774S}) and therefore the visibility of r~modes would be at maximum at intermediate inclinations. However, given the large uncertainties associated with the derived stellar inclination angles (Fig. \ref{fig:area_hump_incl}), it is difficult to fully understand how the star's inclination influences the visibility of r~modes.

\begin{figure}

	\includegraphics[width=\columnwidth]{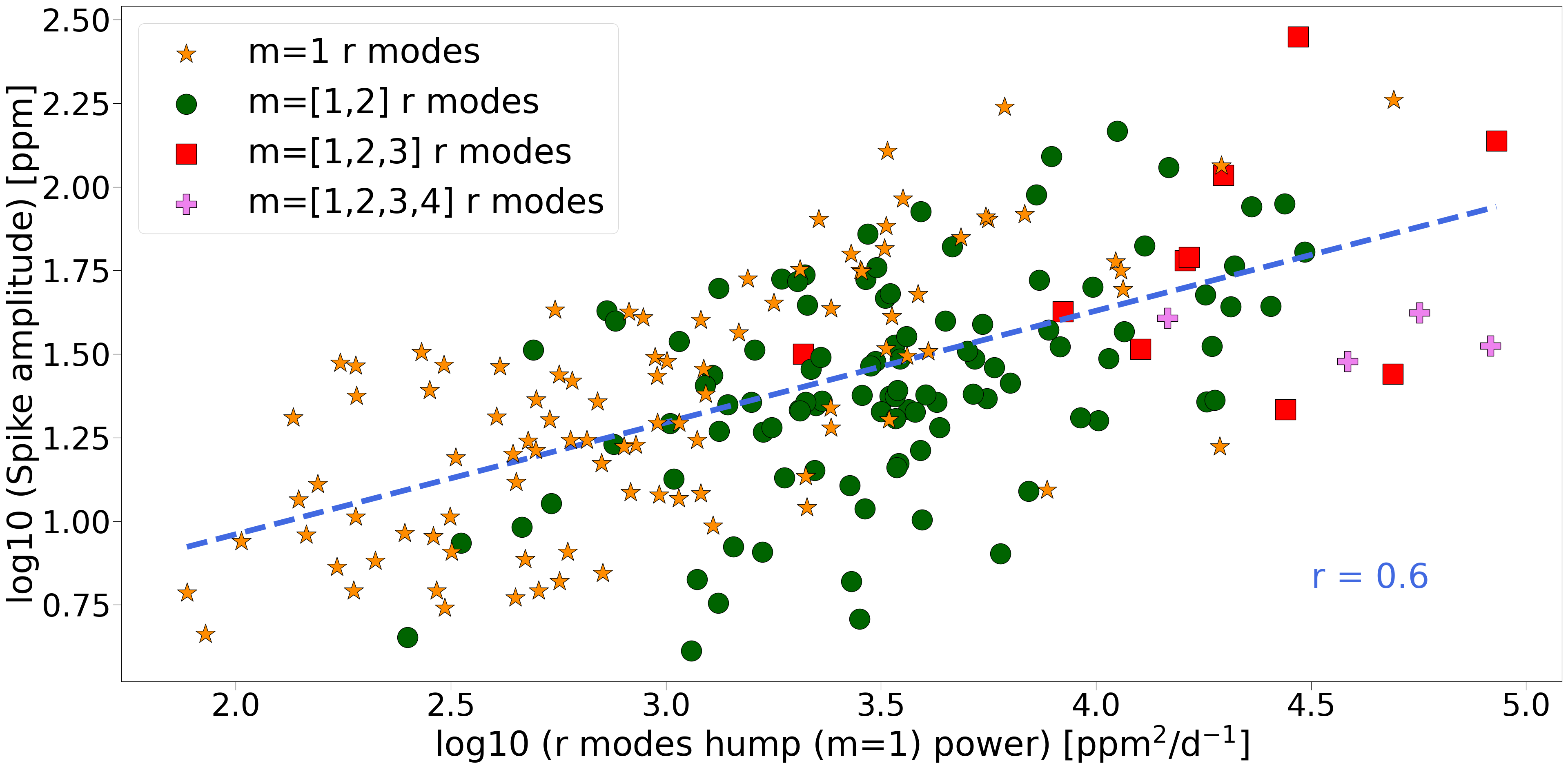}
     \includegraphics[width=\columnwidth]{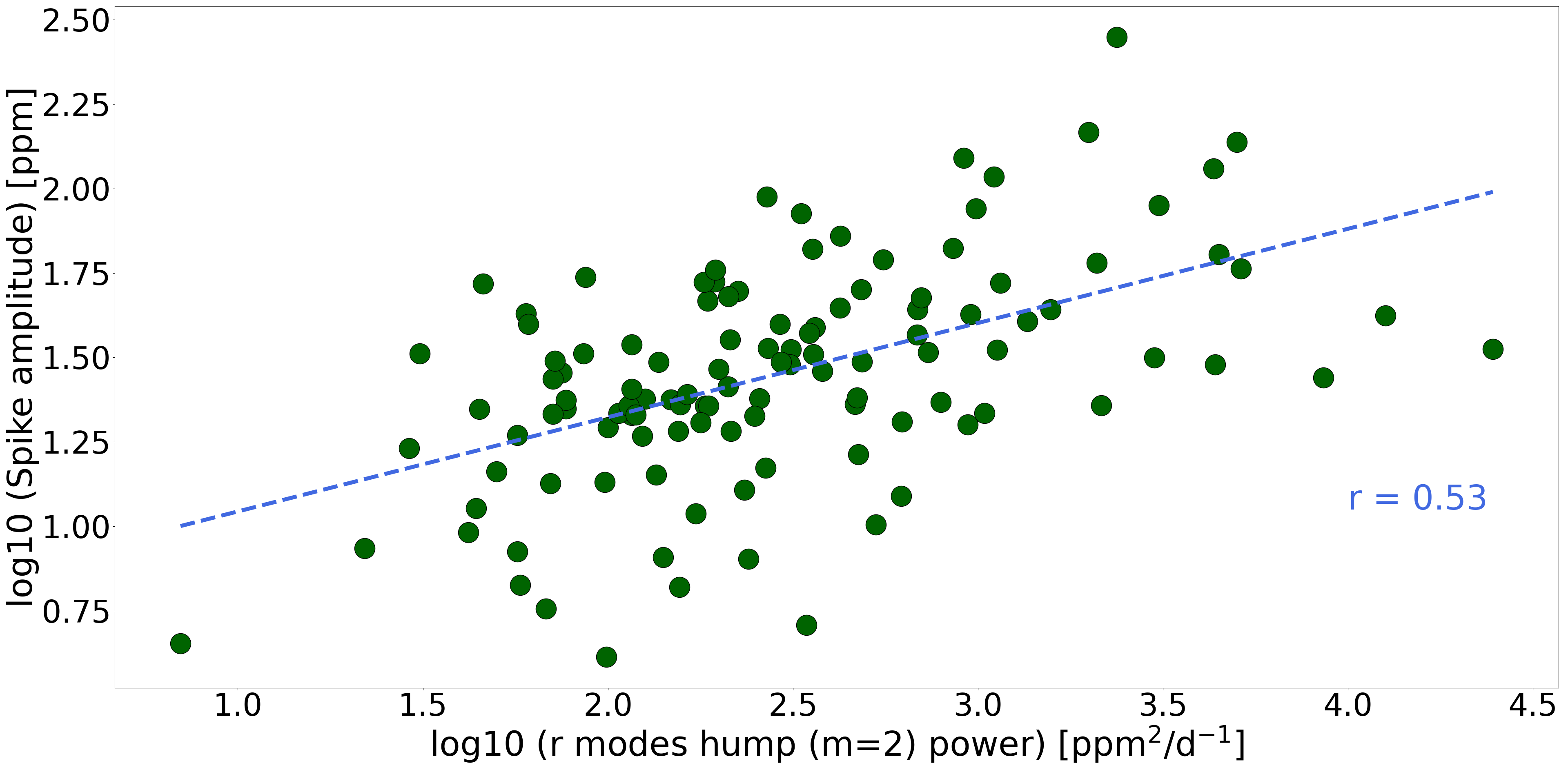}
    \caption{\textit{Upper panel}: Moderate-strong correlation (r = 0.6) between power in the $m=1$ r~modes hump and amplitude of the main spike. Both spike amplitude and power in the $m=1$ r~modes hump can be consulted in Table \ref{tab:spike_hump_stellar_param}. The color coding is similar to that of Fig.\,\ref{fig:area_hump_incl}, where, for example, m = [1,2,3] r~modes means that a star exhibits all $m=1$, $m=2$ and $m=3$ r~mode humps. 
    \textit{Lower panel}: Moderate correlation (r = 0.53) between power in the $m=2$ r~modes hump and amplitude of the main spike. The power in the $m=2$ r~modes hump can be consulted in Table \ref{tab:2nd_hump_spike}.}
    \label{fig:area_hump_spike_ampl}
\end{figure}

The upper panel of Fig. \ref{fig:area_hump_spike_ampl} shows a moderate correlation between the r~modes $m=1$ hump and the amplitude of the spike, suggesting that if stellar spots are responsible for the spike, then a stronger deviation of the flow, induced by magnetic fields, results in higher power r~modes (a similar correlation is also found between the r~modes $m=2$ hump and the amplitude of the spike - lower panel of Fig. \ref{fig:area_hump_spike_ampl}). In the case of OsC modes this would indicate a stronger coupling between the resonantly excited g~modes by the convective core and the mechanically excited r~modes humps. In addition, the color-scheme of Fig. \ref{fig:area_hump_spike_ampl} (upper panel) shows that stars with humps at high azimuthal orders, have more power in the r~modes hump at $m=1$. This means that there is a certain lower power limit for $m=1$ r~modes required in order for $m \geq 2$ to be excited and/or visible.

\begin{figure*}

	\includegraphics[width=\textwidth]{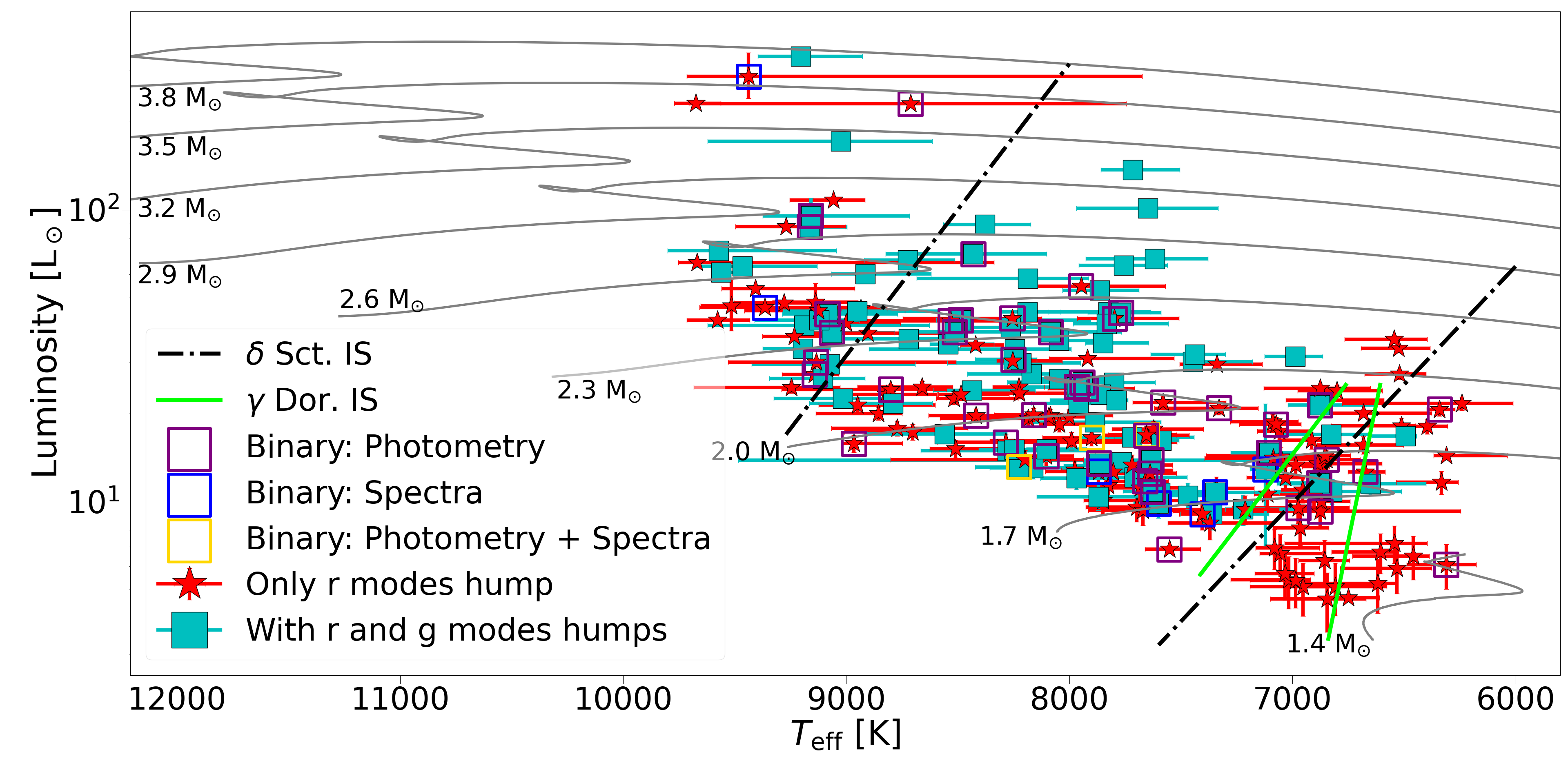}
    \caption{HR diagram displaying 212 stars from our sample. Cyan squares and red stars symbols depict star with and  without g~modes humps, respectively. Some targets are flagged as possible binaries (Binary: Photometry - binarity evidence in photometric data - see example in Fig. \ref{fig:example_hump_spike}; Binary: Spectra - see Table \ref{tab:rv_results_multiple_epochs}). In the background, Warszaw-New\,Jersey evolutionary tracks ($Z=0.012$, \citealt{2004A&A...417..751A}) are displayed for guidance only. The observed $\delta$ Sct. instability strip from \citealt{2019MNRAS.485.2380M} and the theoretical $\gamma$ Dor. instability strip (mixing length parameter,$\alpha = 2$) from \citealt{2005A&A...435..927D} are also displayed. KIC\,5458880 and KIC\,5980337 are not shown here as no luminosity values could be found in literature. Both these targets are flagged as binaries.}
    \label{fig:HRD_hump_before_after}
\end{figure*}

\begin{figure}

	\includegraphics[width=\columnwidth]{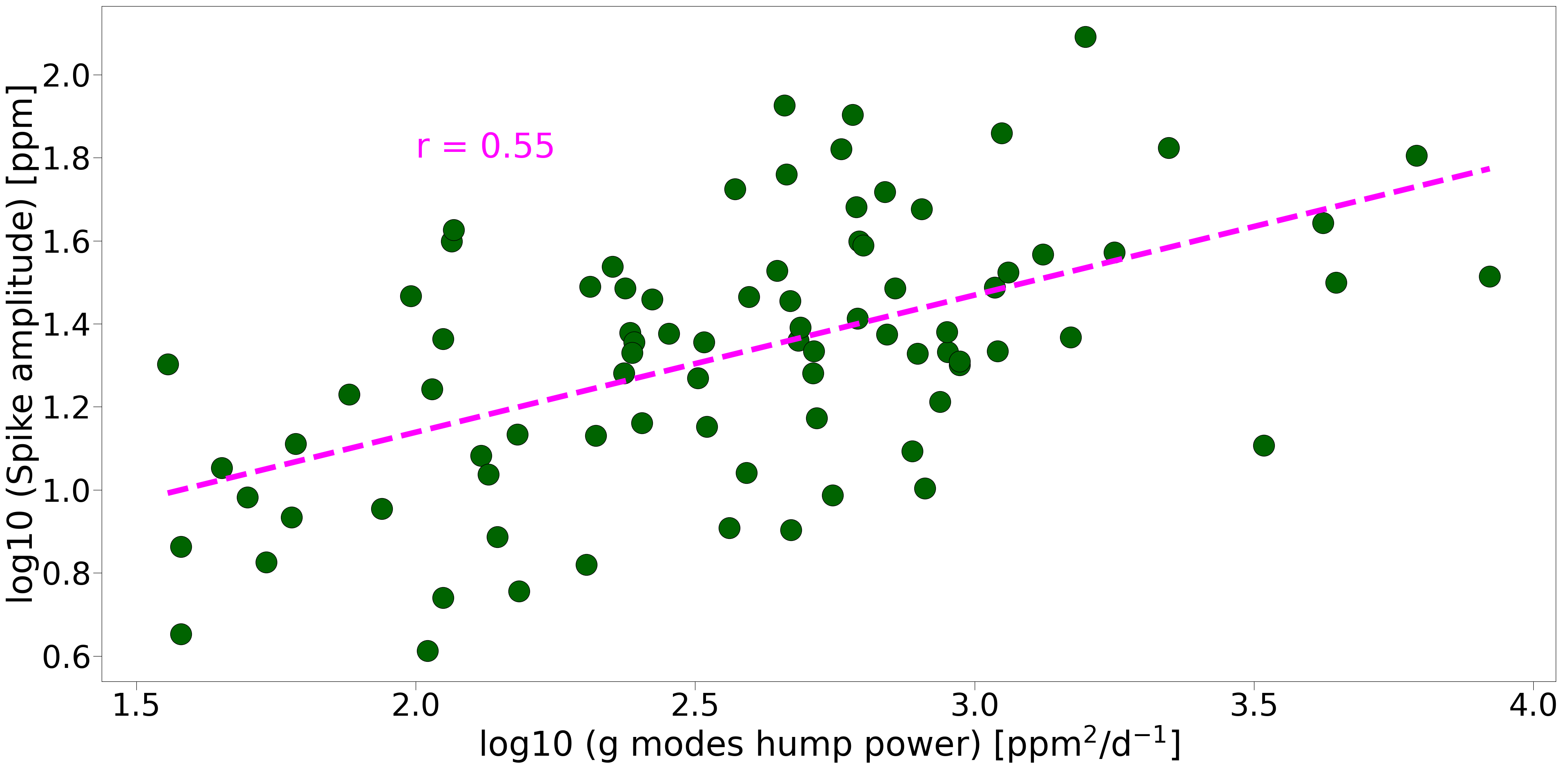}
    \caption{Moderate correlation (r = 0.55) between power in the g~modes humps the amplitudes of the spikes ($m=1$).}
    \label{fig:area_hump1a_spike_ampl}
\end{figure}

In Fig.\,\ref{fig:HRD_hump_before_after}, the stars which possess a g~modes hump are depicted in an HR diagram together with stars that do not. It is interesting that g~modes humps appear in stars hotter than the red edge of the observed $\delta$ Scuti instability strip from \citet{2019MNRAS.485.2380M} (we note that the lack of stars with 
$T_{\rm eff} >10,000K$, is a selection bias, as discussed in 
\citealt{Henriksen23}). If the spike is caused by magnetic stellar spots, this may indicate an interaction between the g~modes in the cooler stars and magnetic fields, leading to possible damping. However, the correlation between g modes humps and the spike amplitudes shown in Fig.\,\ref{fig:area_hump1a_spike_ampl} suggest a different or a more complex explanation.

\begin{figure}
    \includegraphics[width=\columnwidth]{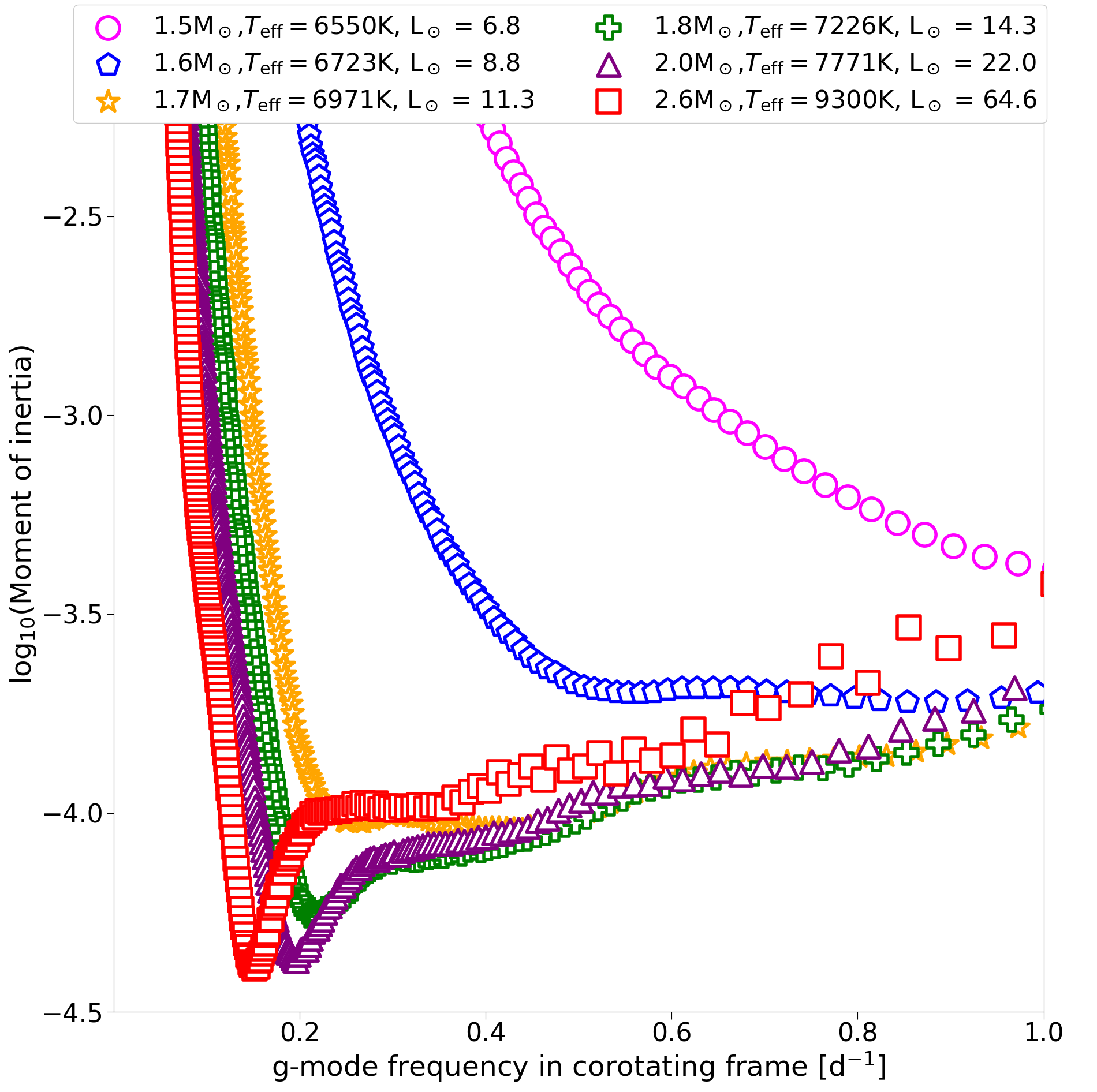}
    \caption{The relationship between the moment of inertia (as defined in equation (\ref{eq:inert})) and the dipole prograde g-mode oscillation frequency in the co-rotating frame for selected stellar models. The models represented here are evolved main-sequence models with central helium abundances of approximately 0.8, which is consistent with the fact that the \textit{hump and spike} stars are also somewhat evolved, as demonstrated in Fig.\,\ref{fig:HRD_hump_before_after}.}
    \label{fig:inert_gmodes}
\end{figure}

\begin{figure}

	\includegraphics[width=\columnwidth]{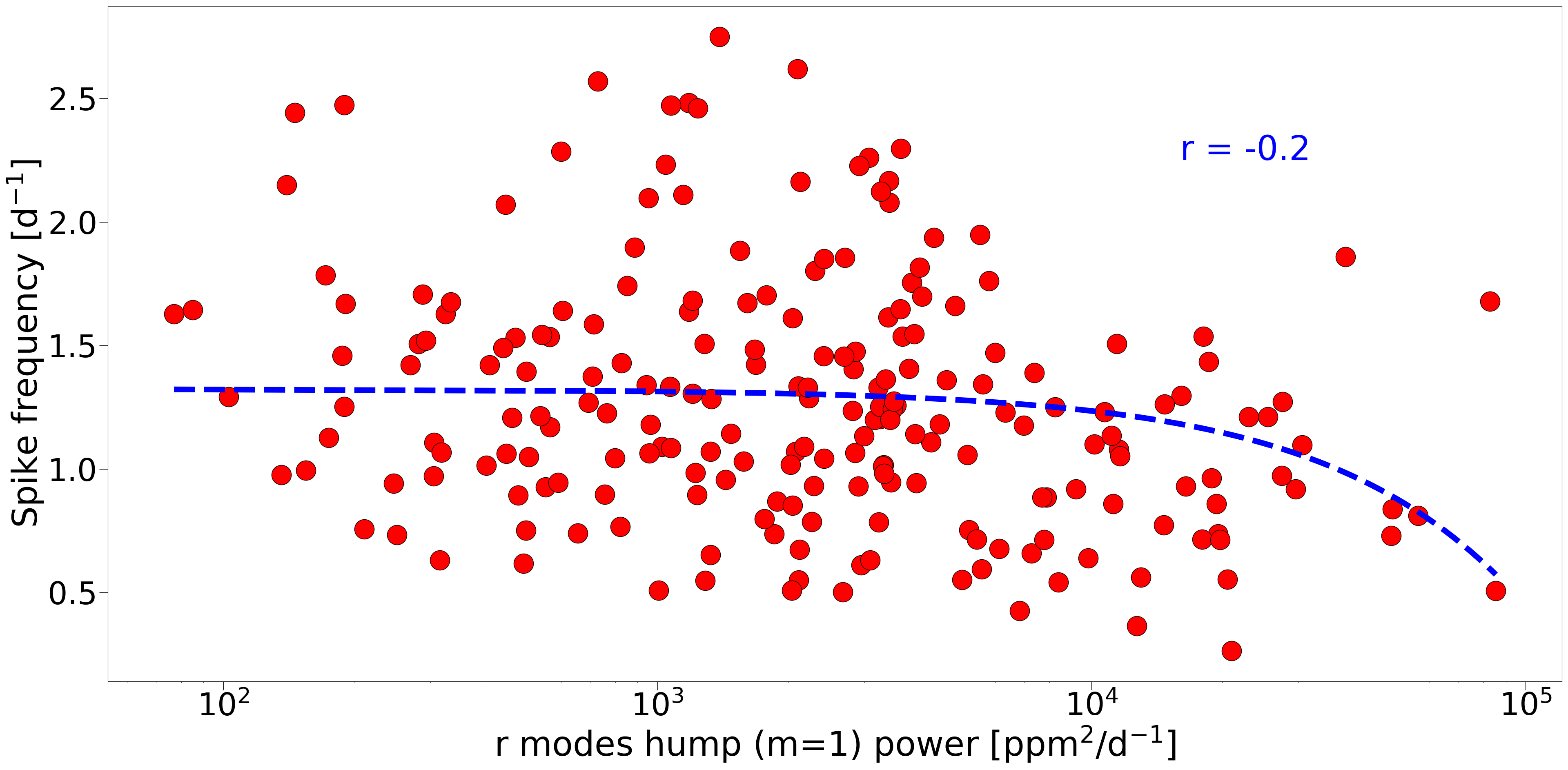}
 	\includegraphics[width=\columnwidth]{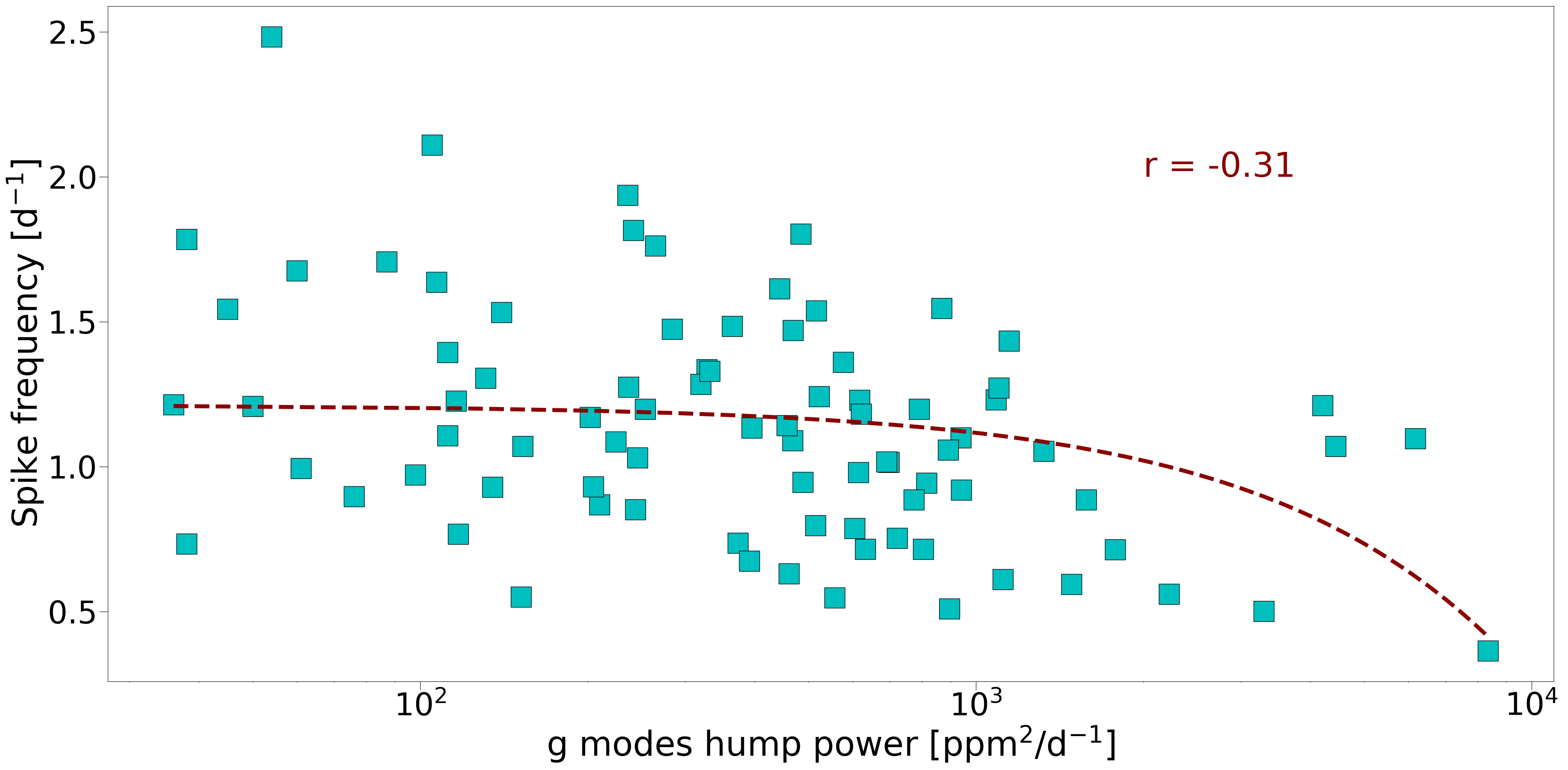}
    \caption{\textit{Upper panel}: Weak anti-correlation (r=-0.2) between power in the r~modes humps and the spike frequencies. \textit{Lower panel}: Weak anti-correlation (r=-0.31) between power in the g~modes humps and the spike frequencies.}
    \label{fig:area_hump_spike_freq}
\end{figure}

The relationship between the moment of inertia and oscillation frequency of g modes provides insights into the possible cause of missing g modes humps, as well as the reason why the frequencies of g modes humps are separated from the spike frequency by approximately 0.1 to 0.25 d$^{-1}$. The moment of inertia of an oscillation mode $k$ is defined as

\begin{equation}
    I_k = \int_0^M \mathbf{\xi}_k\cdot\mathbf{\xi}_k r^2dM_r
    \label{eq:inert}
\end{equation}
where $\mathbf{\xi}_k$ is the displacement vector of the mode. Fig.\,\ref{fig:inert_gmodes} shows the moment of inertia versus frequency of dipole prograde g modes in selected stellar models, in which $I_k$ is normalized by $4\pi R^5\rho_c$ with $R$ and $\rho_c$ being the stellar radius and the central density, respectively.
The g mode frequencies and corresponding eigenfunctions are obtained using the traditional approximation of rotation (TAR), as described by \citet{Saio2018gammdor}. 
For our calculations, we have adopted a rotation frequency of 1\,d$^{-1}$ as a representative rotation (or spike) frequency (Fig.\,\ref{fig:area_hump_spike_freq}). We have adopted the stellar models obtained in \citet{2021MNRAS.502.5856S} where the MESA code was used (v.7184; \citealt{Paxton2011, Paxton2013, Paxton2015}) with an initial chemical composition of (X,Z)=(0.72,0.014). The convective core boundary was determined by the   Schwarzschild criterion, without including a core overshooting parameter. The elemental diffusion was activated for a smooth Brunt-V\"ais\"al\"a frequency. In order to avoid the settling of too much helium, the radiation turbulence was also activated. Rotational deformation was not taken into account. 
The oscillation energy of a mode is given as ${\frac{1}{2}}\sigma_k^2I_k$ with the angular frequency of the mode $\sigma_k$. Therefore, modes with a smaller moment of inertia should be excited to larger amplitudes under some mechanical disturbances than modes with a larger moment of inertia. A g-mode hump would therefore be formed around a minimum in the $I_k$ versus frequency relation.

Notably, Fig.\,\ref{fig:inert_gmodes} reveals dips in the frequency range of approximately 0.15 to 0.25 d$^{-1}$ in models with $M > 1.7\,M_\odot$, which correspond to the frequencies of g mode humps. These dips in $I_k$ coincide with the frequencies of the humps due to the fact that a prograde dipole g mode frequency in the co-rotating frame is equivalent to the frequency difference from the rotation frequency in the observer's frame. Conversely, no such dips appear in less massive models with $M \le 1.7,M_\odot$, which corresponds to the absence of g-mode humps in the cooler side of the instability strip in the HR diagram (see Fig.\,\ref{fig:HRD_hump_before_after}).

\begin{figure}
    \centering
    \includegraphics[width=\columnwidth]{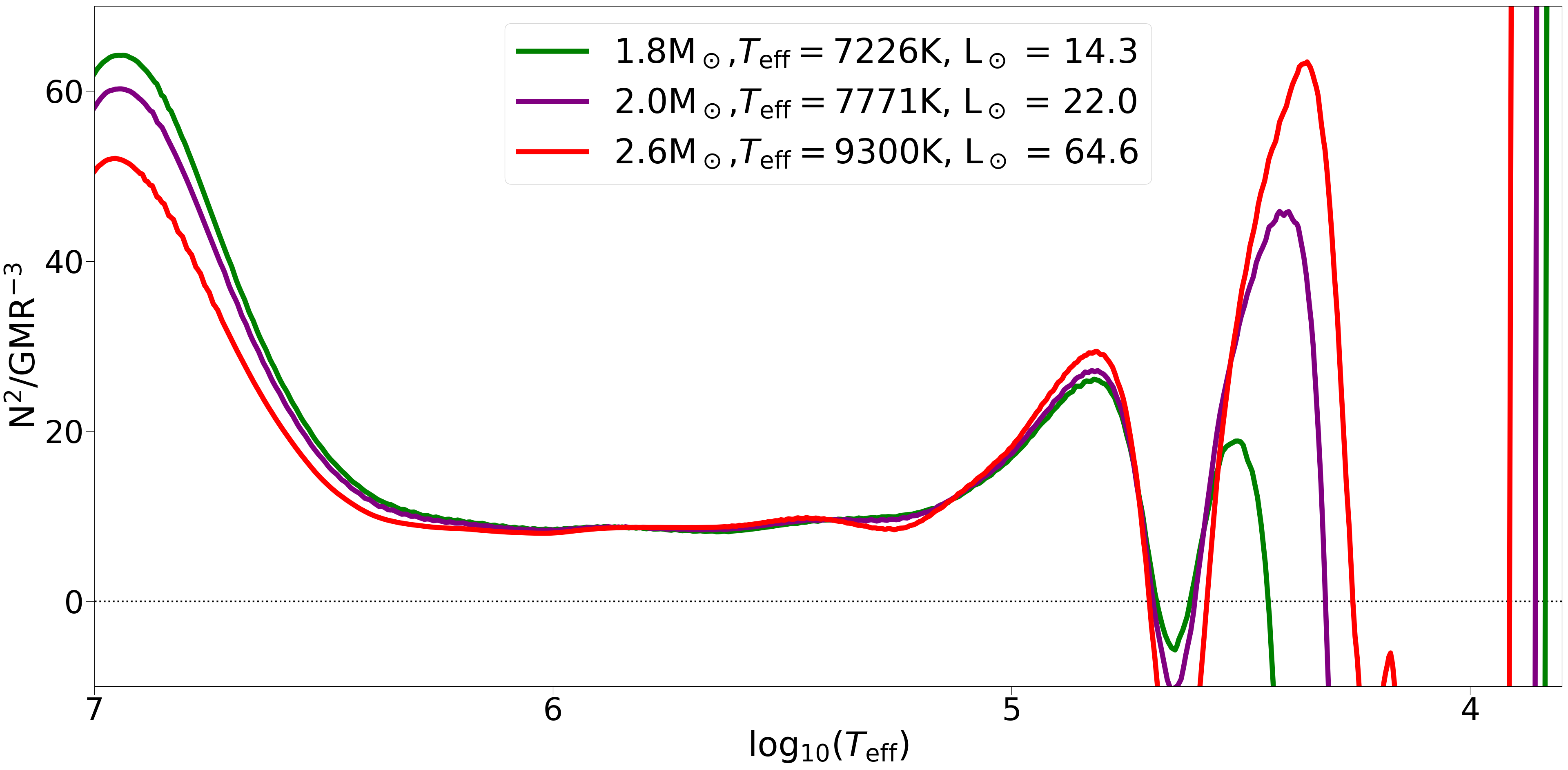}
    \includegraphics[width=\columnwidth]{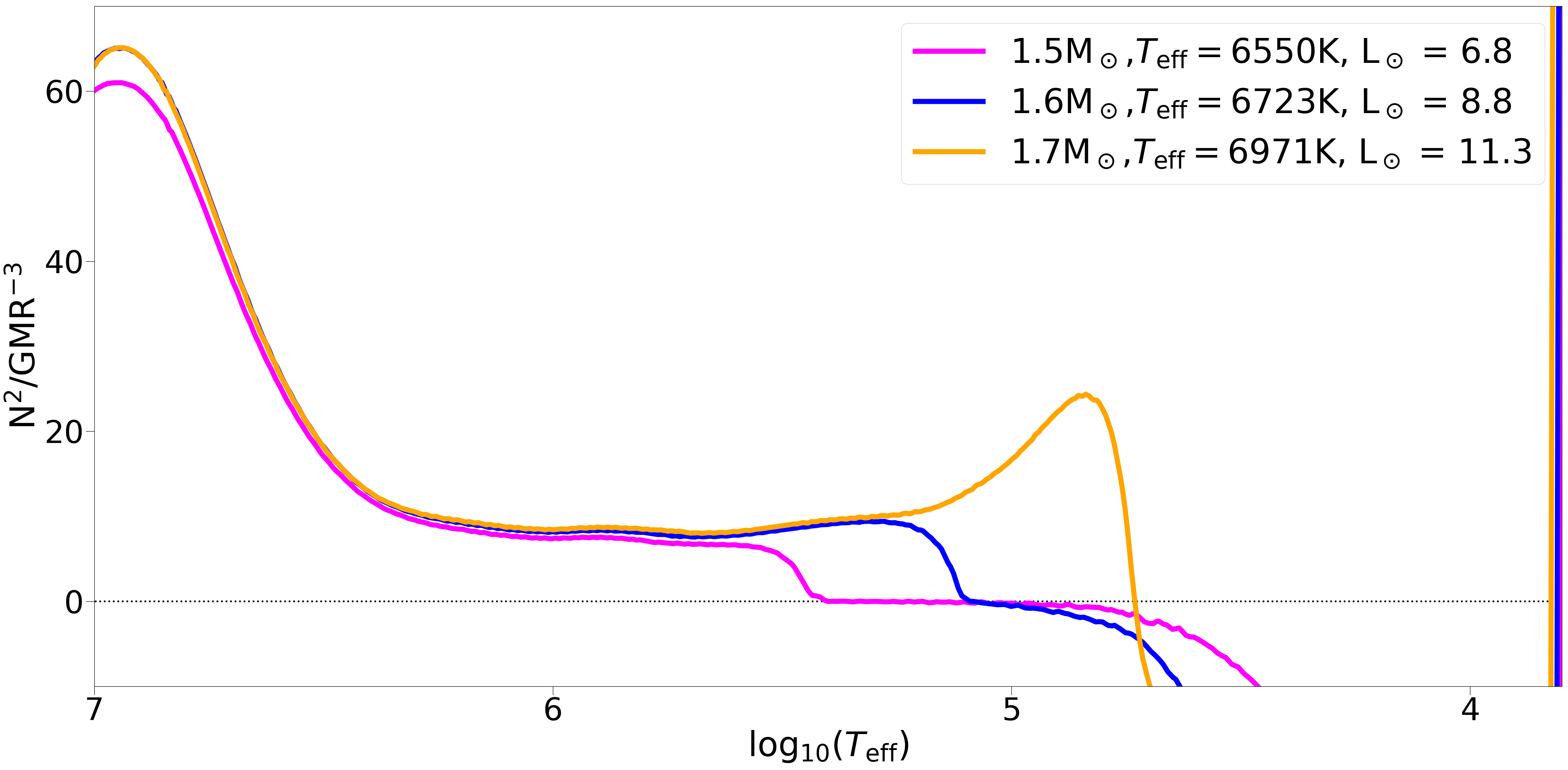}

    \caption{The relationship between the squared normalized Brunt-V\"ais\"al\"a frequency and the logarithmic interior temperature for each model in Fig.\,\ref{fig:inert_gmodes}. The right side of the figure is closest to the surface. \textit{Upper panel}: Data for more massive stars that exhibit a thin radiative zone between the convection zones caused by H/HeI and the HeII ionization. \textit{Lower panel}: Data for less massive models that lack such a zone. }
    \label{fig:propd}
\end{figure}

To investigate the origin of the differences in g-mode properties among the models displayed in Fig.\,\ref{fig:inert_gmodes}, we show the squared Brunt-V\"ais\"al\"a frequency, $N^2$, as a function of the interior temperature for each model in Fig.\,\ref{fig:propd}. The upper panel shows $N^2$ for $1.8, 2.0$ and $2.6\,M_\odot$ models, which exhibit dips in the $I_k$ versus frequency relation around the frequency range between approximately $0.15\,{\rm d^{-1}}$and $0.2\,{\rm d^{-1}}$. In contrast, the lower panel displays $N^2$ for $1.5, 1.6,$ and $1.7\,M_\odot$ models that lack these dips. It is noteworthy that each model in the upper panel has a thin radiative zone between the convection zones associated with H/HeI ionization ($4\la \log_{10}T \la 4.2$) and the HeII ionization at $\log_{10}T\sim 4.6$. Furthermore, the eigenfunction of a g mode at the minimum of the moment of inertia (Fig.\,\ref{fig:inert_gmodes}) has its first node (from the surface) at the bottom boundary of the radiative zone. This observation indicates that the presence of a thin radiative zone between H/HeI and HeII ionization zones is a necessary condition for the narrow minima in the moment of inertia versus g-mode frequency and hence for the occurrence of g-mode humps. In other words, the absence of the thin radiative zone could explain the lack of g-mode humps in stars located in the cooler side of the $\delta$ Sct instability strip (Fig.\,\ref{fig:HRD_hump_before_after}). We note however that this conclusion is based on theoretical models, which contain a simplified picture of the physics in the outer stellar envelope. In reality, convection and overshooting are a complex phenomenon that cannot be fully described in 1D models.

\begin{figure}

	\includegraphics[width=\columnwidth]{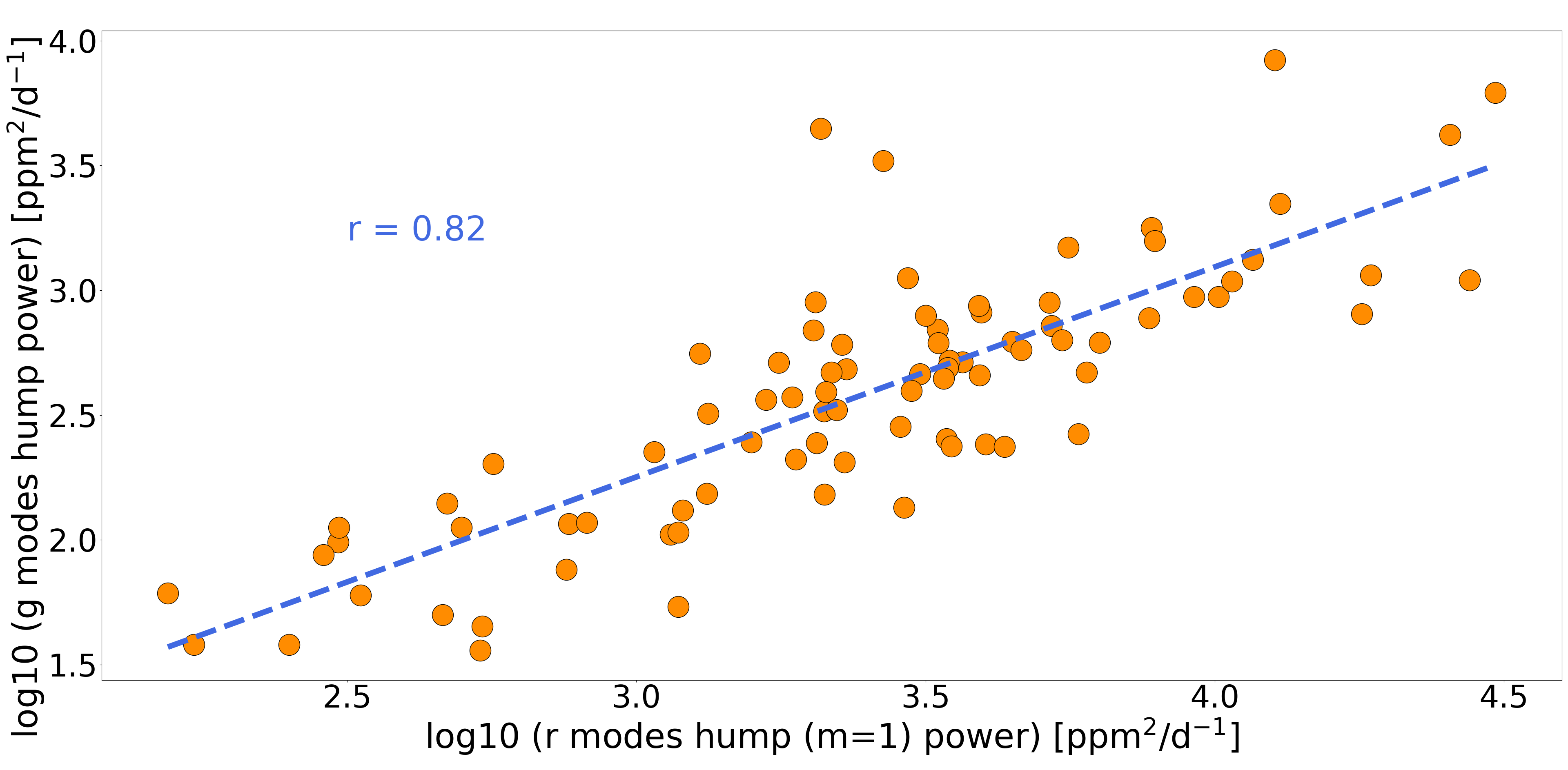}
    \caption{Strong correlation (r=0.82) between the power in the g and r~modes humps ($m=1$).}
    \label{fig:area_hump1_area_hump1a}
\end{figure}

Fig. \ref{fig:area_hump1a_spike_ampl} shows a moderate correlation between the g~modes hump and the amplitude of the spike, suggesting that these are also mechanically excited, by stellar spots or OsC modes. 
Furthermore, the strong correlation between the r~modes ($m=1$) and the g~modes humps (Fig. \ref{fig:area_hump1_area_hump1a}) clearly indicates a common mechanism generating the two types of modes.

\begin{figure}

	\includegraphics[width=\columnwidth]{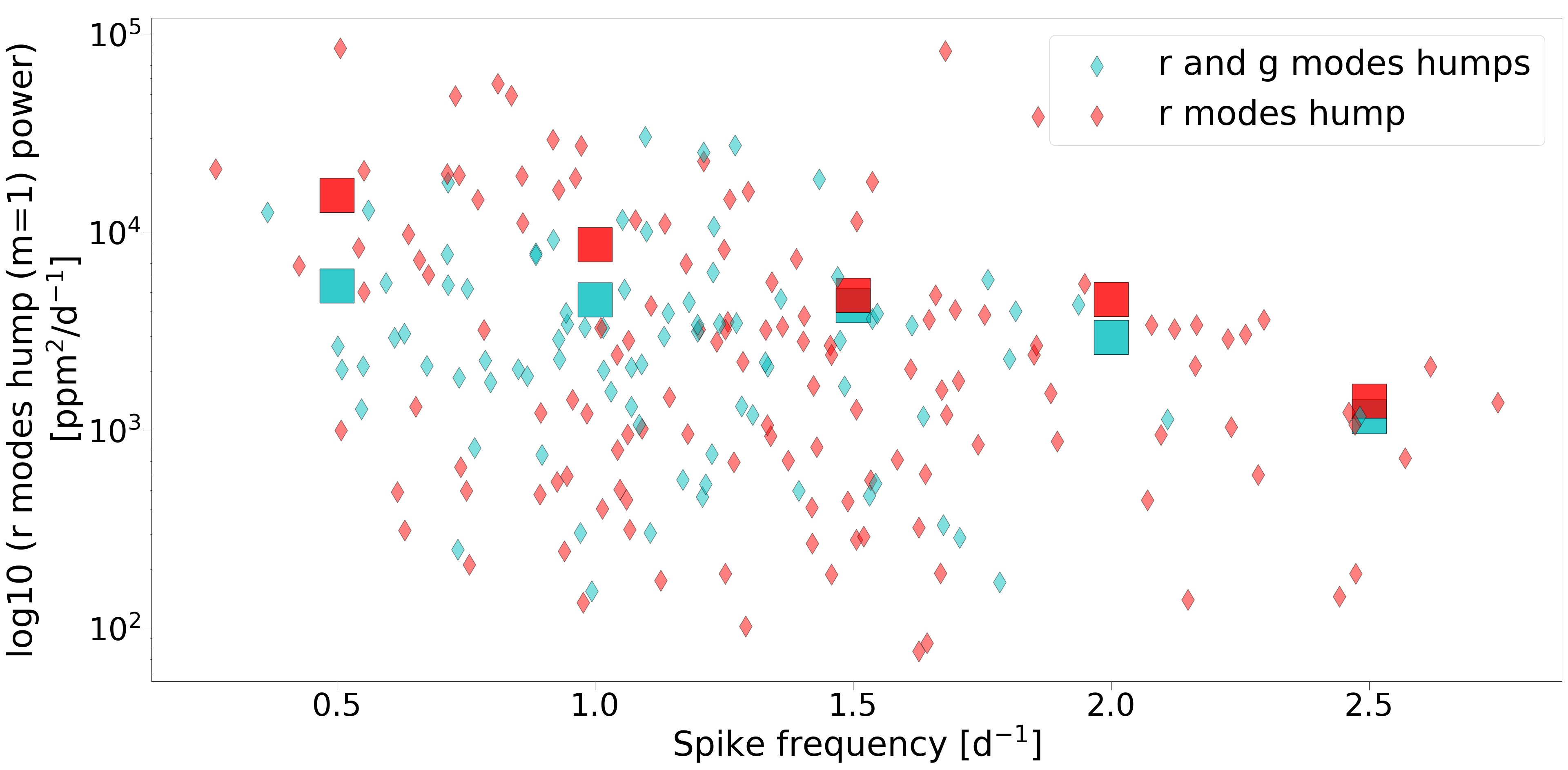}
    \caption{Rotation frequency (spike frequency) as a function of the r~modes hump ($m=1$) power. Here we distinguish between stars that exhibit the g~modes hump and those that do not. Small markers in background are the values for each stars. Large markers indicate the average values of r~modes hump ($m=1$) power for stars that with spike frequencies within the following ranges [0.25-0.75 $\rm\,d^{-1}$], [0.75-1.25$\rm\,d^{-1}$], [1.25-1.75$\rm\,d^{-1}$], [1.75-2.25$\rm\,d^{-1}$] and [2.25-2.75$\rm\,d^{-1}$]. }
    \label{fig:area_hump_before_frot_bin}
\end{figure}

The weak correlation between the rotation frequency (spike frequency), the r and the g~modes humps (upper panel and lower panel, respectively, of Fig. \ref{fig:area_hump_spike_freq}) suggests that the rotation rate does not play an essential role in the excitation of these modes. Furthermore, stars that possess the g~modes humps and those that do not are not distinct concerning their rotation rates (see Fig. \ref{fig:area_hump_before_frot_bin}). However, given the low amplitude of the g~modes hump signal, it cannot be excluded that these features may be present in all stars, but buried in the noise. 

\begin{figure}

	\includegraphics[width=\columnwidth]{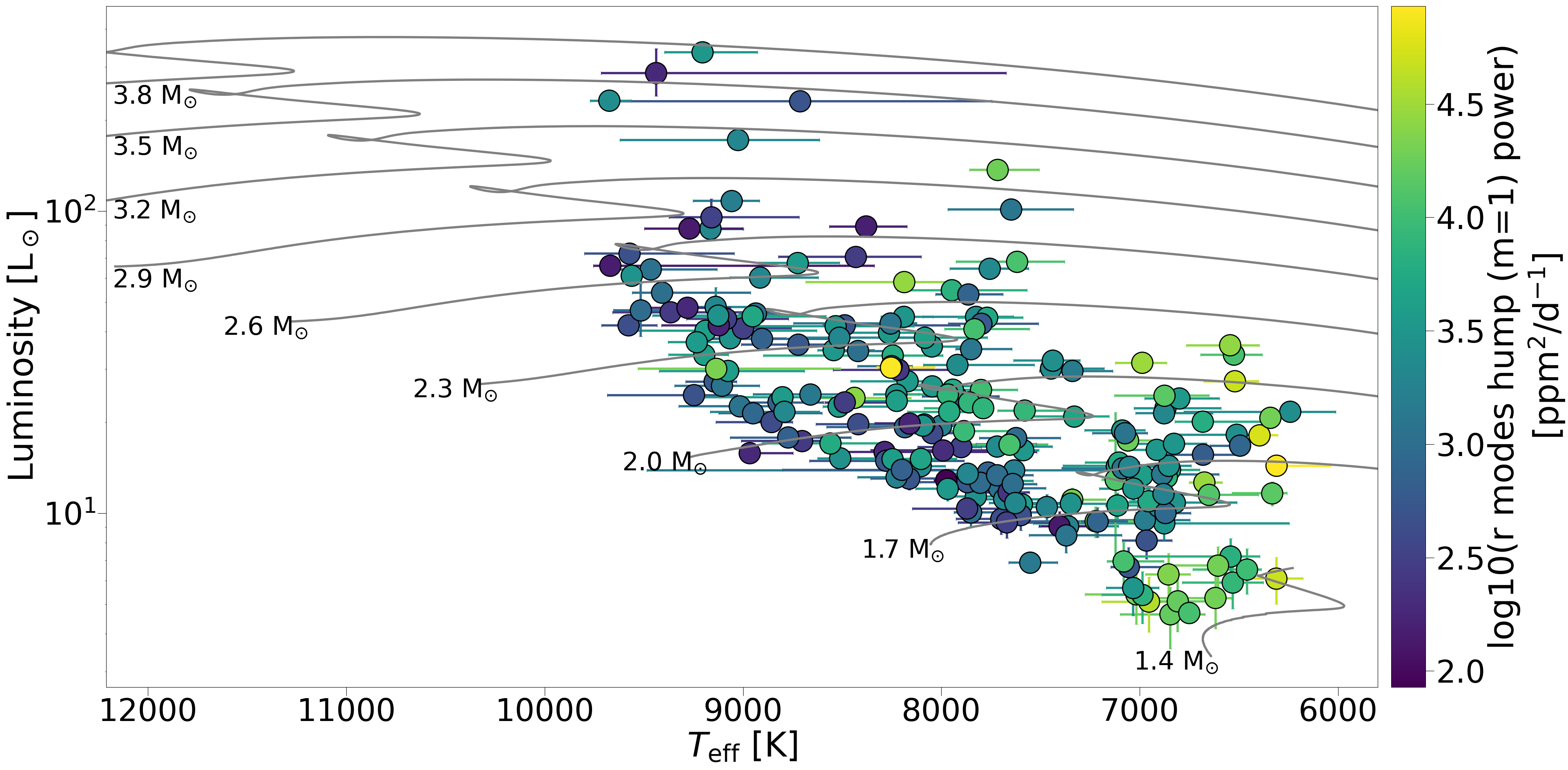}
    \caption{HR diagram with the symbol colour indicating the power in r~modes $m=1$ humps. In the background, Warszaw-New\,Jersey evolutionary tracks ($Z=0.012$, \citealt{2004A&A...417..751A}) are displayed for guidance only. KIC\,5458880 and KIC\,5980337 are not shown here as no luminosity values could be found in literature. Both these targets are flagged as binaries. }
    \label{fig:HRD_area_hump}
\end{figure}

Similar to fig. 17 in \citet{Henriksen23}, where the spike amplitude correlates with the stellar mass, Fig. \ref{fig:HRD_area_hump} shows a similar but less obvious trend, of the r~modes $m=1$ power decreasing with stellar mass. This is not surprising given the correlation displayed in Fig. \ref{fig:area_hump_spike_ampl} (upper panel). However, we note that possible companions could contaminate the $T_{\rm eff}$ and luminosity values. It could be that the trend seen in Fig. \ref{fig:HRD_area_hump} may become stronger, if more precise and reliable stellar parameters would be determined in the future. Furthermore, a more in-depth focus on the binarity occurrence in \textit{hump and spike} stars would be a valuable endeavor, but more spectroscopic data are required. Given the faintness of the targets, a significant amount of data is required. We note that some stars may have been flagged as binaries in literature, however their orbital periods turned out to be the spike frequencies, which clearly suggests erroneous identification.

\section{Conclusions}
\label{sec:conc}
This work is a continuation of the ensemble analysis from \citet{Henriksen23}, where a detailed investigation of the \textit{spike} in \textit{hump and spike} stars was conducted. Here we concentrate on the \textit{hump} signal and use both \textit{Kepler} photometry and ground-based spectroscopy to characterize 214 stars.  
We measure the power in r~modes humps and find that a significant number of stars display an additional hump at slightly higher frequencies than the spike, for which we also determine the power. We use spectroscopic data to derive RV measurements and investigate possible binarity. Out of the 22 stars, 11 are consistent with RV shifts due to a companion. Furthermore, we determine \textit{v}\,sini\,\textit{i} values for 28 stars and find literature values for additional 61 stars. Together with the rotational velocities derived from the rotational frequencies and stellar radii, we estimate the stellar inclination angle for 89 stars. 
Based on our results, we conclude the following:

\begin{enumerate}
\item We find a strong correlation between the r and the g~modes humps and spike amplitudes, suggesting that either stellar spots of OsC modes could indeed induce both types of oscillations mechanically (Fig. \ref{fig:area_hump_spike_ampl} and \ref{fig:area_hump1a_spike_ampl}).
\item Rotation does not seem to play an important role, as the spike frequency (due to rotational modulation in the core or at the surface) only weakly correlates the power in either the r and g~modes humps (Fig. \ref{fig:area_hump_spike_freq}). This is consistent with \citet{Henriksen23}, who report no correlation between the spike amplitudes and frequencies.

\item The larger the power in the r~mode hump ($m=1$), the more likely it is for higher azimuthal r~modes to be excited and/or visible (upper panel Fig. \ref{fig:area_hump_spike_ampl}).
\item Stars cooler than the observed red $\delta$ Scuti instability strip show no g~mode humps, suggesting that the presence of a thin radiative zone between H/HeI and HeII ionization zones is a necessary condition for the occurrence of g-modes humps (Fig. \ref{fig:HRD_hump_before_after}).

\end{enumerate}

We assess that at this stage, neither the stellar magnetic spots nor OsC modes can be excluded as the mechanism for the spike mechanically inducing the r and g~modes humps. We conclude that detailed modelling of selected stars needs to be conducted; however, the period spacings of high-order r~modes cannot be retrieved for these stars. We show that r~modes with higher amplitudes would have to be observed for almost a century in order to be resolved, which is unfeasible. To distinguish between the stellar spots or OsC scenarios, we recommend spectro-polarimetric observations of bright \textit{hump and spike} stars that could be identified in the continuous viewing zone of the TESS mission. Another open question that should be addressed in future work is why not all A and F stars that show rotational modulation are \textit{hump and spike} stars? Comparing these two groups of stars may shed more light on this phenomenon.  
 
\section*{Acknowledgements}

Based on observations made with the Nordic Optical Telescope, owned in collaboration by the University of Turku and Aarhus University, and operated jointly by Aarhus University, the University of Turku and the University of Oslo, representing Denmark, Finland and Norway, the University of Iceland and Stockholm University at the Observatorio del Roque de los Muchachos, La Palma, Spain, of the Instituto de Astrofisica de Canarias under programs 64-013 and 65-854.

Based on observations made with the Mercator Telescope, operated on the island of La Palma by the Flemish Community, at the Spanish Observatorio del Roque de los Muchachos of the Instituto de Astrofísica de Canarias. 

Based on observations obtained with the HERMES spectrograph, which is supported by the Research Foundation - Flanders (FWO), Belgium, the Research Council of KU Leuven, Belgium, the Fonds National de la Recherche Scientifique (F.R.S.-FNRS), Belgium, the Royal Observatory of Belgium, the Observatoire de Genève, Switzerland and the Thüringer Landessternwarte Tautenburg, Germany. 

Based on observations made with the Hertzsprung SONG telescope operated on the Spanish Observatorio del Teide on the island of Tenerife
by the Aarhus and Copenhagen Universities and by the Instituto de Astrofísica de Canarias.

This work was supported by a research grant (00028173) from VILLUM FONDEN. Funding for the Stellar Astrophysics Centre is provided by The Danish National Research Foundation (Grant agreement no.: DNRF106).

This research was supported in part by the National Science Foundation under Grant No. NSF PHY-1748958.

D.M.B. gratefully acknowledges funding from the Research Foundation Flanders (FWO) by means of a senior postdoctoral fellowship (grant agreement No. 1286521N).

TVR gratefully acknowledges support from the Research Foundation Flanders (grant number 12ZB620N) and from the KU Leuven Research Council (grant C16/18/005: PARADISE).

\section*{Data Availability}

The data underlying this article are available in the article and in its online supplementary material.



\bibliographystyle{mnras}
\bibliography{example} 




\appendix

\section{Some extra material}


\begin{table*}
	\caption{Spectroscopic observations analysed in this work - meta data. KIC - \textit{Kepler} input catalogue identification number; Instrument - Spectrograph with which the spectrum was acquired; Mid-time stamp = Midpoint of observation; Exposure time in seconds. Only the first 10 rows are displayed here. The full table is available online; here, the first 10 rows are shown for guidance on content and style. The full version contains 75 lines.}

	\label{tab:spectra_metadata}
	\begin{tabular}{ccccc} 
	\hline
    \hline
    KIC&Instrument&Filename&Mid-time stamp&Exposure time[s]\\
    \hline
    3634487&FIES&FIFf190083.fits&2022-06-20T00:36:29&1200\\
    3634487&FIES&FIFe260039.fits&2022-05-27T00:38:37&1200\\
    3848948&FIES&FIFf190107.fits&2022-06-20T03:55:55&1100\\
    3848948&FIES&FIFe250037.fits&2022-05-26T02:10:21&1100\\
    4066110&FIES&FIEj160096.fits&2021-10-16T21:29:23&1500\\
    4066110&FIES&FIEj160097.fits&2021-10-16T21:55:10&1500\\
    4066110&FIES&FIEj160098.fits&2021-10-16T22:20:57&1500\\
    4572373&HERMES&00413721\_HRF\_OBJ.fits&2012-07-14T01:55:49&2400\\
    4572373&HERMES&00419068\_HRF\_OBJ.fits&2012-09-05T00:43:08&1800\\
    4572373&HERMES&00419069\_HRF\_OBJ.fits&2012-09-05T01:13:58&1800\\
     \hline
 \hline
	\end{tabular}
\end{table*}

\begin{table}
	\caption{\textit{v}\,sin\,\textit{i} values taken from literature and measured in this work, and the estimated inclination values. The full table is available online; here, the first 10 rows are shown for guidance on content and style. The full version contains 87 lines.}
    \setlength{\tabcolsep}{4.1pt} 
	\label{tab:vsini_incl}
	\begin{tabular}{ccccccc} 
    \hline
    \hline
    KIC&\textit{v} sin \textit{i}&$\sigma_{\textit{v} {\rm sin} \textit{i}}$& Ref.& \textit{i} & $\sigma_{\textit{i}}$ & \\
    & [$\mathrm{km\,s^{-1}}$]& [$\mathrm{km\,s^{-1}}$]& \textit{v} sin \textit{i}&[deg] &[deg] \\
    \hline
    3459226 & 73 & 5 & 1 & 61 & 8 \\
    3634487 &175 & 14 & 0 & 62 & 8 \\
    3766112 & 332 & 33 & 2 & 75 & 7 \\
    3848948 &98 & 12 & 0 & 53 & 9 \\
    3868032 & 181 & 9 & 3 & 59 & 5 \\
    4059089 & 280 & 72 & 2 & 56 & 15 \\
    4066110 &138 & 19 & 0 & 66 & 11 \\
    4481029 & 136 & 88 & 2 & 32 & 28 \\
    4488313 & 208 & 54 & 2 & 58 & 15 \\
    4567097 & 83 & 5 & 1 & 24 & 2 \\

\hline
\multicolumn{6}{l}{Ref. (\textit{v} sin \textit{i} source): (0) This work; (1) \cite{2021MNRAS.504.5528T};} \\
\multicolumn{6}{l}{(2) \cite{2016AA...594A..39F}; (3) \cite{2020MNRAS.493.4518K};} \\
\multicolumn{6}{l}{(4) \cite{2015MNRAS.450.2764N}; (5) \cite{2011AA...526A.124L};}\\
\multicolumn{6}{l}{(6) \cite{2010AA...517A...3C}; (7) \cite{2020MNRAS.498.2456S};} \\
\multicolumn{6}{l}{(8) \cite{2013MNRAS.431.3685T}, (9) \cite{2016MNRAS.457.3988B}} \\
\multicolumn{6}{l}{(10) \cite{2014MNRAS.445.2446M}
}\\
\multicolumn{6}{l}{\textit{i}, $\sigma_{\textit{i}}$: stellar inclination and its uncertainty. }\\
\hline
	\end{tabular}
\end{table}

\begin{table*}
    \centering
    \setlength{\tabcolsep}{1.9pt} 
    \caption{Extracted spike and hump parameters, and stellar parameters. 
    $R, \, R_{\rm p}, \,\rm R_{\rm m} $: Radius value, upper and lower uncertainties; R ref. = radius source: b = \citealt{2020AJ....159..280B}, g2 = \textit{Gaia} DR2, g3 = \textit{Gaia} DR3,  k = Kepler Input Catalogue v10 \citep{2011AJ....142..112B};
    $f_{\rm rot}, \sigma_{f_{\rm rot}}$: Spike frequency and associated standard deviation;
    $A_{\mathrm{ rot}}$: Spike amplitude; $v_{\rm rot}, \, v_{\rm rot_p}, \, v_{\rm rot_{m}}$: Rotational velocity and associated uncertainties;
    $T_{\rm eff},\,T_{\rm eff_p}, \,T_{\rm eff_m}$: Effective temperature and upper and lower uncertainties - from \textit{Gaia} DR2, except for KIC\,8462852 (taken from \citealt{2016MNRAS.457.3988B}) ; $ L,\,L_{\rm p}, \,L_{\rm m}$: Luminosity and upper and lower uncertainties; L ref. = luminosity source: b = \citealt{2020AJ....159..280B}, b16 = \citealt{2016MNRAS.457.3988B}, g = \textit{Gaia} DR2, m = \citealt{2019MNRAS.485.2380M};  $\rm {Power_{hump}}$ - Power in the r~modes hump;  $f_{\rm hump_l}$ , $f_{\rm hump_r}$ - lower and higher frequency limits of the hump. The full table is available online; here, the first 10 rows are shown for guidance on content and style. The full version contains 215 lines. Note that there are two lines for KIC\,5980337 as there are two \textit{hump and spike} features that could originate from a binary system. No luminosity values are given for KIC\,5458880 and KIC\,5980337.}
    \label{tab:spike_hump_stellar_param}
    \begin{tabular}{ccccccccccccccccccccc}
    \hline
    \hline
    
    KIC &  $R$&$ R_{\rm p}$	&$\rm R_{\rm m} $& $R$& $f_{\rm rot}$  & $\sigma_{f_{\rm rot}}$  & $A_{\rm rot}$ &  $v_{\rm rot}$ &  $v_{\rm rot_p}$ &  $v_{\rm rot_m}$ &  $T_{\rm eff}$	&$T_{\rm eff_p}$ &	   $T_{\rm eff_m}$ & $ L$	&$L_{\rm p}$	&$L_{\rm m}$ &  $L$ &
    ${\rm Power_{hump}}$& $f_{\rm hump_l}$ & $f_{\rm hump_r}$  \\

    & $[{\rm R}_{\odot}$]& [${\rm R}_{\odot}$]&[${\rm R}_{\odot}$] &ref. &[$\mathrm{d^{-1}}$] & [$\mathrm{d^{-1}}$] & [ppm]& [$\mathrm{km\,s^{-1}}$] & [$\mathrm{km\,s^{-1}}$] & [$\mathrm{km\,s^{-1}}$] & [K] & [K] & [K] & [${\rm L}_{\odot}$] &[${\rm L}_{\odot}$]& [${\rm L}_{\odot}$] &ref.& [$\mathrm {ppm^2/d^{-1}}$] & [$\mathrm{d^{-1}}$] & [$\mathrm{d^{-1}}$] \\

    \hline
    1722916&1.5&0.03&0.03&g3&0.553&0.0013&44&42&1&1&7017&260&98&5.4&1.09&1.09&m&20556&0.455&0.530 \\
    1873552&1.6&0.04&0.04&g3&1.236&0.0016&5&100&3&3&7826&121&86&11.4&1.05&1.06&m&2818&1.024&1.221 \\
    2157489&1.9&0.04&0.04&g3&0.737&0.0014&53&71&1&1&7361&140&254&9.1&1.05&1.05&m&1859&0.644&0.728 \\
    2158190&2.8&0.11&0.09&b&1.016&0.0006&24&144&6&5&8048&42&219&35.6&1.05&1.05&m&3317&0.934&1.008 \\
    3002336&2.4&0.05&0.06&g3&1.202&0.0004&76&146&3&4&7115&272&217&14.4&1.06&1.07&m&3254&1.056&1.190 \\
    3222104&2.1&0.07&0.07&g3&1.270&0.0004&23&135&4&4&8821&231&228&23.3&1.08&1.08&m&694&1.095&1.263 \\
    3238627&2.8&0.08&0.08&g3&1.537&0.0024&23&218&6&6&7059&98&140&17.4&1.10&1.10&m&18076&1.444&1.530 \\
    3240406&1.9&0.04&0.04&g3&1.071&0.0013&6&103&2&2&7865&161&201&13.3&1.06&1.06&m&1324&0.972&1.066 \\
    3337124&1.8&0.14&0.03&g2&1.108&0.0004&23&101&8&2&7835&61&298&12.8&1.05&1.05&m&4268&0.962&1.101 \\
    3440710&1.6&0.04&0.04&g3&1.250&0.0021&33&101&3&3&6531&256&154&5.9&1.06&1.06&m&8248&1.075&1.242 \\
        \hline
    \end{tabular}

\end{table*}

\begin{table}
    \centering
    \caption{Wavelength limits of the sections into which the HERMES spectra were split.}
    \label{tab:hermes_sections}
    \begin{tabular}{c|c|c}
    \hline
    \hline
        & \multicolumn{2}{c}{Wavelength limits} \\ 
        \cline{2-3}
    Section & blue & red \\
            & [\r{A}] & [\r{A}] \\
            \hline
    1 & 4000 & 4225 \\
    2 & 4225 & 4450 \\
    3 & 4450 & 4550 \\
    4 & 4550 & 4650 \\
    5 & 4650 & 4750 \\
    6 & 4750 & 5034 \\
    7 & 5087 & 5200 \\
    8 & 5200 & 5414 \\
    9 & 5462 & 5683 \\
    10 & 5754 & 5870 \\
    11 & 6000 & 6275 \\
    12 & 6361 & 6432 \\
    13 & 6600 & 6800 \\
    \hline
    \hline
    \end{tabular}

\end{table}

\begin{table*}
	\caption{Spectral lines used for \textit{v}\,sin\,\textit{i} determination. The resulting \textit{v}\,sin\,\textit{i} values are listed next to the corresponding spectral lines (denoted by the species, ionization state and wavelength) for each star (identified with KIC ID). In some cases, an additional identifier for the used line is the échelle order, as some lines can be found in two adjacent orders that overlap (e.g., order 27 and 28  = o27, o28). The final \textit{v}\,sin\,\textit{i} values are listed in Table \ref{tab:vsini_kics} as the average values obtained from all spectral lines. }

	\label{tab:vsini_lines_kic_1}
	\begin{tabular}{ccccccccc} 
		\hline
        \hline
    KIC& Spectral line  &  \textit{v}\,sin\,\textit{i} & KIC& Spectral line  &  \textit{v}\,sin\,\textit{i} & KIC&Spectral line  &  \textit{v}\,sin\,\textit{i}  \\
    & [\r{A}] & [${\rm km\ s^{-1}}$] & & [\r{A}] & [${\rm km\ s^{-1}}$]  && [\r{A}] & [${\rm km\ s^{-1}}$] \\
    
    \hline
    3634487 &  Fe II 4508.29 (o27) & 177 & 5938266 & Fe I 6065.48  & 112 & 7939835 & Ti II 4501.27 & 25  \\
      & Fe II 4508.29 (o28) & 177 & &Fe I 6078.49 & 109 & & Fe II 4508.29 (o27)  & 23  \\
      & Ti II 4571.97 & 154 & & Fe I 6393.61 & 114 & &Fe II 4508.29 (o28)  & 21 \\
      & Ca I 6439.07 & 192 &  & Ca I 6439.07  & 76 & & Ti II 4571.97 & 23\\
      \cline{1-6}

      3848948 &  Fe II 4508.29 (o27) & 97 &5980337 & Fe I 6055.99& 77 &  & Fe II 4923.93 & 17\\
      & Fe II 4508.29 (o28) & 112 & & Fe I 6065.48 &111 & & Fe I 5862.35 & 18 \\
      & Ti II 4571.97 & 80 & & Ca I 6122.22 & 67  &  & Fe I 5934.65 & 38\\
      & Ca I 6439.07 & 104 &  & Ca I 6439.07 & 91  &  & Fe I 6020.17 & 26\\
      \cline{1-6}

       4066110 &  Fe I 6065.48 & 140 &  6039039& Fe I 5862.35 &75 &  & Fe I 6024.05 & 19\\
    &  Fe I 6078.49 & 161 &  & Fe I 6055.99 & 81  & & Fe I 6055.99 & 34\\
    &  Ca I 6439.07  & 114 &  & Fe I 6065.48 & 85  &   & Fe I 6065.48 & 24\\
    
     \cline{1-3} 
        4661914 & Fe I 5633.98 & 84  & &Fe I 6078.49 & 79 &  & Ca I 6122.22 & 18\\
    & Fe I 5905.69  &94   & & Ca I 6122.22 & 66  & & Fe I  6393.61 & 27\\
    \cline{4-6}
    & Fe I 6065.48  &62 &  6192566 & Fe II 4508.29 & 48 & & Ca I 6439.07 & 17\\
    \cline{7-9}

    &  Fe I 6078.49 & 100 & & Ti II 4571.97 & 47 & 7959579&  Fe I 6065.48 & 120  \\

    \cline{1-3}

    4856799 &Fe I 4404.76 & 186 & & Fe I 6055.99 & 56   & &Fe I 6078.49 &  120 \\
     & Fe II 4508.29 (o27) & 177 & & Fe I 6065.48 & 65  &  & Fe I  6393.61 & 109\\
     & Fe II 4508.29 (o28) & 182 & & Ca I 6122.22 & 32  & & Ca I 6439.07 & 120 \\
     \cline{7-9}
     & Ti II 4571.97 & 173 & & Fe I  6393.61 & 59 &  8037519 & Si II 6371.35 & 177 \\  
     \cline{4-6}
     & Fe II 4923.93 & 147 &  6610433 & Fe II 4508.29 (o27) & 110 & & Mn II 4481.13/33 & 196  \\

    \cline{1-3} \cline{7-9}
 
    5024410 &Fe I 4454.39 & 7 &  & Fe II 4508.29 (o28) & 126 & 8783760  & Ti II 4779.99 & 105  \\
           & Fe I 4508.29 & 8 & & Fe I 5633.98 & 119 &  & Fe I 6065.48 & 148   \\
           & Ti I 4518.02 & 7 &  & Ca I 6439.07  & 117 &  & Fe I 6078.49  & 144\\
    \cline{4-6}
           & Ti I 4548.76 & 5 & 6974705 & Mn II 4481.13/33& 246 & &  Ca I 6122.22 & 124\\
    \cline{4-6}
           & Ti II 4563.76 & 7 & 7116117 & Fe I 5633.98& 72 &  &  Ca I 6439.07 & 116\\
    \cline{7-9}
           & Fe II 4576.34 & 6 & & Fe I 5905.69 & 80 & 8846809 & Fe II 4508.29 (o27) & 162\\
           & Ti II 4583.41 &  8 & &  Fe I 6065.48 & 54 &  &Ti II 4571.97 &161\\
           & Cr II 4588.2 & 7 & &  Fe I 6078.49 & 76 & &Fe I 6065.48 &140 \\
    \cline{4-6}\cline{7-9}
           & Fe I 4602.94 &11 & 7131587& Fe II 4508.29 (o27) & 40 & 9163520& Fe II 4508.29& 37 \\
           & Cr I 4652.15 & 6 &  & Fe II 4508.29 (o28) & 44 &  & Fe I 5576.09 & 42  \\
           & Mn I 4739.09 &10 &  & Fe I 5862.35&  52 & & Fe I 6055.99 & 47\\
           & Fe I 4741.53 &6 & & Fe II 5987.07& 55 & & Fe I 6065.48 & 40 \\
           & Fe II 4923.93 & 12 & & Fe II 6020.17&  34 &  & Fe I 6078.49 & 35  \\
           & Fe I 5090.77 & 9  & & Fe I 6024.05& 48 & & Fe I  6393.61 & 44 \\
    \cline{7-9}
           & Fe I 5141.74 & 5 & & Fe I 6027.05& 58 & 9273647&  Fe I 6065.48 & 121\\
           & Fe I 5151.91 & 6 & & Fe I 6055.99& 44  & &Fe I 6078.49 &  123  \\
           & Ti I 5210.39 & 6 & & Fe I 6065.48&  50 & & Fe I  6393.61 & 112     \\
           & Fe I 5364.86 & 10 & &  Fe I 6078.49& 64 &  & Ca I 6439.07 & 108\\
    \cline{7-9}
           & Fe I 5398.28 & 12  & & Si II 6371.36 (o64)&  47 & 9519698 &Fe I 4404.76 & 182\\
           & Fe I 5862.35 & 6  & & Si II 6371.36 (o65)& 54 & & Fe II 4508.29 (o27) & 177 \\
           & Fe I 6024.05 & 4 & & Fe I 6430.86 & 41  & & Fe II 4508.29 (o28) & 177  \\
           & Fe I 6065.48 & 6 & & Ca I 6439.07 & 39 &   & Ti II 4571.97 & 177 \\
           & Ca I 6122.22 &7 &  & Ca I 6471.67&  60 & & Fe II 4923.93 & 164\\
     \cline{4-6} \cline{7-9}
           & Fe I 6252.57 &5 & 7842339 & Fe I 4404.76 & 93 & 10068389 & Ti II 4501.27 & 124 \\
           & Fe I 6393.61 & 6 & & Fe II 4508.29 & 101 & & Fe II 4508.29 (o27) & 133  \\
    \cline{1-3}
        5456027 & Fe II 4508.29 (o27) & 158&  & Ca I 6439.07& 119 & & Fe II 4508.29 (o28) & 127 \\
    \cline{4-6}
      & Fe II 4508.29 (o28) & 156 &  &  &  & & Ti II 4571.97 & 127   \\
      & Ti II 4571.97 & 157 & &  &  & & Ca I 6439.07 & 150 \\
    \cline{7-9}
      & Ca I 6439.07 & 171  & & &  & 10810140 & Ti II 4501.27 & 136 \\
     \cline{1-3}
      5566579 & Fe II 4515.34 & 169 & & &  && Fe II 4508.29 (o27) & 133 \\
      & Ti II 4571.97 & 128  & & & &  &Fe II 4508.29 (o28) & 129\\
      & Fe II 4923.93 & 125  & & &  && Ti II 4571.97 & 128\\
    \cline{1-3}
    5903499 & Mn II 4481.13/33& 254  & & &&  & Ca I 6439.07 & 122\\

\hline
\hline
	\end{tabular}
\end{table*}


\begin{table}
	\caption{Information about the second harmonic of the spike and $m=2$ r~modes hump.  $f, \sigma_{f}$: Frequency of second harmonic of the spike and associated standard deviation;
    $A_{\mathrm{ rot}}$: Amplitude of second harmonic of the spike; $\rm {Power_{hump}}$ - Power in the $m=2$ r~modes hump;  $f_{\rm hump_l}$ , $f_{\rm hump_r}$ - lower and higher frequency limits of the $m=2$ r~modes hump. Only the first 10 rows are displayed here. The full table is available online; here, the first 10 rows are shown for guidance on content and style. The full version contains 112 lines.}

	\label{tab:2nd_hump_spike}
	\begin{tabular}{ccccccc} 
    \hline
    \hline
    KIC & $f$ & $\sigma_{f}$ & A & $\mathrm{Power_{hump}}$&  $f_{\rm hump_l}$ & $f_{\rm hump_r}$\\
    &$[\rm d^{-1}]$ & $[\rm d^{-1}]$ &$\rm{[ppm]} $ & $\rm{[ppm^2/d^{-1}]}$ &$[\rm d^{-1}]$  &$[\rm d^{-1}]$ \\
    \hline
1722916&1.105&0.0014&10.0&686&0.958&1.016 \\
2157489&1.474&0.0004&7.8&194&1.326&1.467 \\
2158190&2.032&0.0006&6.3&148&1.908&2.017 \\
3238627&3.077&0.0005&18.5&2152&2.976&3.060 \\
3240406&2.141&0.0008&4.4&68&1.999&2.096 \\
3337124&2.216&0.0004&3.4&183&1.911&2.200 \\
3440710&2.501&0.0021&19.3&1128&2.319&2.487 \\
3766112&4.463&0.0007&4.8&70&4.355&4.459 \\
3848948&2.574&0.0006&2.0&45&2.403&2.568 \\
3868032&3.358&0.0019&6.7&24636&0.754&0.819 \\

 \hline
	\end{tabular}
\end{table}


\bsp	
\label{lastpage}
\end{document}